\documentclass{vldb}
\usepackage{graphicx}
\usepackage{balance}  
\usepackage[ruled, vlined, linesnumbered]{algorithm2e}
\usepackage{multirow,multicol,enumitem,epsfig,url,array,makecell,balance,tabularx,cite}
\usepackage{xcolor}
\usepackage{comment}
\usepackage{subcaption}
\usepackage{tablefootnote}

\vldbTitle{}
\vldbAuthors{}
\vldbDOI{}
\vldbVolume{}
\vldbNumber{}
\vldbYear{2020}

\newtheorem{theorem} {Theorem}
\newtheorem{lemma} {Lemma}

\newdef{definition}{Definition}

\newcommand{\hide}[1]{}

\newcommand{\todo}[1]{{\color{red}{\bf [#1]}}}

\newcommand{\revise}[1]{{{#1}}}

\def\ours{$\mathsf{SimPush}$\xspace}

\def\leftpush{Source\text{-}Push\xspace}
\def\leftpushnospace{Source\text{-}Push\xspace}

\def\revspread{Reverse\text{-}Push\xspace}
\def\revspreadnospace{Reverse\text{-}Push\xspace}
\def\nevermeetcorr{Last\text{-}Meeting\text{ }Correction\xspace}
\def\nevermeetcorrnospace{Last\text{-}Meeting\text{ }Correction\xspace}

\def\probesim{$\mathsf{ProbeSim}$\xspace}
\def\prsim{$\mathsf{PRSim}$\xspace}
\def\sling{$\mathsf{SLING}$\xspace}
\def\reads{$\mathsf{READS}$\xspace}
\def\tsf{$\mathsf{TSF}$\xspace}
\def\topsim{$\mathsf{TopSim}$\xspace}

\def\sim{s}
\def\simap{\tilde{s}}
\def\simapleft{s'}
\def\simapleftgamma{s^{+}}
\def\src{u}
\def\Tsrc{G_u}
\def\Tdst{G_v}
\def\Aset{A_u}
\def\Asetl{\Aset^{(\ell)}}
\def\Asetli{\Aset^{(\ell+i)}}
\def\Asetlj{\Aset^{(\ell+j)}}

\def\c{c}
\def\eps{\epsilon}
\def\epsp{\epsilon_{h}}
\def\sqrtc{\sqrt{c}}
\def\hl{h^{(\ell)}}
\def\hzero{h^{(0)}}
\def\hlnext{h^{(\ell+1)}}

\def\hzeroap{\tilde{h}^{(0)}}
\def\hlap{\tilde{h}^{(\ell)}}

\def\hrlap{\hat{h}^{(\ell)}}

\def\inN{\mathcal{I}}
\def\inNTu{\mathcal{I}^{T}}
\def\outN{\mathcal{O}}
\def\outNTu{\mathcal{O}^{T}}
\def\inD{d_{\mathcal{I}}}
\def\inDTu{d^{T}_{\mathcal{I}}}
\def\outD{d_{\mathcal{O}}}
\def\corr{\gamma}
\def\corrl{\gamma^{(\ell)}}

\def\meeti{\rho^{(i)}}
\def\meetone{\rho^{(1)}}
\def\meetj{\rho^{(j)}}
\def\meet{\rho}

\def\lml{\kappa^{(\ell)}}

\def\hinT{\tilde{h}}
\def\hinTi{\tilde{h}^{(i)}}

\def\deltaL{\Delta{l}}
\def\resi{r}
\def\resil{r^{(\ell)}}
\def\resilprime{r^{(\ell^\prime)}}
\def\resilminusone{r^{(\ell-1)}}
\def\resilprimeminusone{r^{(\ell^\prime-1)}}
\def\resilzero{r^{(0)}}

\def\wl{w}

\def\wli{w_i}
\def\wlj{w_j}

\def\pf{\delta}

\def\numwalkshalf{{\log \frac{1}{(1-\sqrtc)\epsp\delta}/\epsp^2}}
\def\numWalks{{2\cdot}\numwalkshalf}
\def\complexleftpush{{m\log\frac{1}{\eps} + \log\frac{1}{\eps\delta}/\eps^2}}
\def\complexnummeeting{{1/\eps}}
\def\complexpushintree{{m\log\frac{1}{\eps}/\eps}}
\def\complexlastmeetsingle{{1/\eps^2}}
\def\complexlastmeetall{{m\log\frac{1}{\eps}/\eps + 1/\eps^3}}
\def\complexrevpush{{m\log\frac{1}{\eps}}}
\def\complexours{{m\log\frac{1}{\eps}/\eps + \log\frac{1}{\eps\delta}/\eps^2 + 1/\eps^3}}

\def\header{\vspace{1.5mm} \noindent}

\begin{document}


\title{
Realtime Index-Free Single Source SimRank Processing on Web-Scale Graphs
}



%
%
%
%

\numberofauthors{1}
\author{
%
%
\alignauthor
Jieming Shi$^\dagger$, Tianyuan Jin$^\ddagger$, Renchi Yang$^\ast$, Xiaokui Xiao$^\dagger$, Yin Yang$^\S$\\
       \affaddr{$^\dagger$School of Computing, National University of Singapore, Singapore}\\
       \affaddr{$^\ddagger$University of Science and Technology of China, Hefei, China}\\
       \affaddr{$^\ast$School of Computer Science and Engineering, Nanyang Technological University, Singapore}\\
       \affaddr{$^\S$College of Science and Engineering, Hamad Bin Khalifa University, Qatar}\\
       \email{$^\dagger$\{shijm, xkxiao\}@nus.edu.sg, $^\ddagger$tianyuan1044@gmail.com,\\
       $^\ast$yang0461@e.ntu.edu.sg,
       $^\S$yyang@hbku.edu.qa}
}


\maketitle

\sloppy

\begin{abstract}

Given a graph $G$ and a node $u \in G$, a single source SimRank query evaluates the similarity between $u$ and every node $v \in G$. Existing approaches to single source SimRank computation incur either long query response time, or expensive pre-computation, which needs to be performed again whenever the graph $G$ changes. Consequently, to our knowledge none of them is ideal for scenarios in which (i) query processing must be done in realtime, and (ii) the underlying graph $G$ is massive, with frequent updates. 

Motivated by this, we propose \ours, a novel algorithm that answers single source SimRank queries without any pre-computation, and at the same time achieves significantly higher query processing speed than even the fastest known index-based solutions. Further, \ours provides rigorous result quality guarantees, and its  high performance does not rely on any strong assumption of the underlying graph. Specifically, compared to existing methods, \ours employs a radically different algorithmic design that focuses on (i) identifying a small number of nodes relevant to the query, and subsequently (ii) computing statistics and performing residue push from these nodes only.

We prove the correctness of \ours, analyze its time complexity, and compare its asymptotic performance with that of existing methods. Meanwhile, we evaluate the practical performance of \ours through extensive experiments on \revise{9 real datasets}. The results demonstrate that \ours consistently outperforms all existing solutions, often by over an order of magnitude. In particular, on a commodity machine, \ours answers a single source SimRank query on a web graph containing over 133 million nodes and 5.4 billion edges in under \revise{62 milliseconds, with 0.00035 empirical error, while the fastest index-based competitor needs 1.18 seconds.}

\end{abstract}

\section{Introduction} \label{sec:intro}

\begin{small}
\begin{table*}[!t]
\renewcommand{\arraystretch}{1.2}
\centering
\caption{Comparison of single-source SimRank algorithms with error tolerance $\epsilon$ and failure probability $\delta$} \label{tbl:algos}
\vspace{-2mm}
\begin{tabular}{|l|c|c|c|}
\hline
\multicolumn{1}{|c|}{{\bf Algorithm}} &
\multicolumn{1}{c|}{{\bf Query Time}} &
\multicolumn{1}{c|}{{\bf Index Size}} &
\multicolumn{1}{c|}{{\bf Preprocessing Time}}\\
\hline
\ours & $O(\complexours)$ & - & -\\
\hline
\tsf \cite{tsfShaoC0LX15} & $O\left(n\log{\frac{n}{\delta}}/\epsilon^2\right)$ & $O\left(n\log{\frac{n}{\delta}}/\epsilon^2\right)$ & $O\left(n\log{\frac{n}{\delta}}/\epsilon^2\right)$ \\
\hline
\reads \cite{readsJiangFWW17} & $O\left(n\log{\frac{n}{\delta}}/\epsilon^2\right)$ & $O\left(n\log{\frac{n}{\delta}}/\epsilon^2\right)$ & $O\left(n\log{\frac{n}{\delta}}/\epsilon^2\right)$ \\
\hline
\probesim \cite{probsimLiuZHWXZL17} & $O\left(n\log{\frac{n}{\delta}}/\epsilon^2\right)$ & - & -\\
\hline
\sling \cite{slingTianX16} & $O(n/\epsilon)$ & $O(n/\epsilon)$ & $O\left(m/\epsilon+n\log{\frac{n}{\delta}}/\epsilon^2\right)$\\
\hline
\prsim \cite{prsimWeiHX0LDW19}\footnotemark[1] & $O\left(n\log{\frac{n}{\delta}}/\epsilon^2\right)$ & $O(\min{\{n/\epsilon,m\}})$ & $O(m/\epsilon)$ \\
\hline
\end{tabular}
\vspace{-2mm}
\end{table*}
\end{small}

SimRank is a popular similarity measure between nodes in a graph, with numerous potential applications, \textit{e.g.}, in recommendation systems \cite{nguyen2015evaluation}, schema matching \cite{melnik2002similarity}, spam detection \cite{benczur2006link}, and graph mining \cite{jin2011axiomatic,liben2007link,ShiMWC14}. The main idea of SimRank is that two nodes that are referenced by many similar nodes are themselves similar to each other. For instance, in a social network, two key opinion leaders who are followed by similar fans are expected to be similar in some way, \textit{e.g.}, sharing similar political positions or life experiences. Formally, given a graph $G$ and nodes $u, v \in G$, the SimRank value $s(u, v)$ between $u$ and $v$ is defined as follows:
\begin{equation*}
s(u,v)=
\begin{cases}
1, & \textrm{if } u=v \\
\frac{c}{|\inN(u)|\cdot|\inN(v)|}\sum\limits_{u'\in \inN(u)}{\sum\limits_{v'\in \inN(v)}{s(u',v')}}, & \textrm{otherwise}.
\end{cases}
\end{equation*}
where $\inN(u)$ and $\inN(v)$ are the sets of in-neighbors of $u$ and $v$, respectively, and $c \in [0, 1]$ is a decay factor commonly fixed to a constant, \textit{e.g.}, $c=0.6$ \cite{slingTianX16,prsimWeiHX0LDW19,probsimLiuZHWXZL17}.

This paper focuses on single-source SimRank processing, which takes as input a node $u\in G$, and computes the SimRank $s(u, v)$ between $u$ and every node $v \in G$. This can be applied, for example, in a search engine that retrieves web pages similar to a given one, or in a social networking site that recommends new connections to a user.  We focus on \textit{online} scenarios, in which (i) query execution needs to be done in realtime, and (ii) the underlying graph can change frequently and unpredictably, meaning that query processing must not rely on heavy pre-computions whose results are expensive to update. For large graphs, this problem is highly challenging, since computing SimRank values is immensely expensive: its original definition, presented above, is recursive and requires numerous iterations over the entire graph to converge, which is clearly unscalable. 

Several recent approaches, notably \cite{topsimLeeLY12, slingTianX16, probsimLiuZHWXZL17, prsimWeiHX0LDW19, tsfShaoC0LX15, readsJiangFWW17}, have demonstrated promising results for single source SimRank processing, by solving the approximate version of the problem with rigorous result quality guarantees, as elaborated in Section \ref{sec:prelim}. The majority of these methods, however, require extensive pre-processing to index the input graph $G$; as explained in Section \ref{sec:stateoftheart}, such indexes cannot be easily updated when the underlying graph $G$ changes, meaning that these methods are not suitable for our target scenarios described above. Specifically, the current state of the art for offline single source SimRank is {\sf PRSim} \cite{prsimWeiHX0LDW19}, which achieves efficient query processing with a relatively lightweight index; nevertheless, it is clearly infeasible to rebuild index for every graph update or new query, as shown in our experiments in Section \ref{sec:experiments}. The current best index-free solution is {\sf ProbeSim} \cite{prsimWeiHX0LDW19}, whose query efficiency is far lower than that of {\sf PRSim}. Consequently, {\sf ProbeSim} yields poor response time for large graphs,  adversely affecting user experience.

This paper proposes \ours, a novel index-free solution for approximate single source SimRank processing that achieves significantly higher performance compared to all existing solutions (including index-based ones with heavy pre-computation), while providing rigorous quality guarantees. This is achieved through a novel algorithmic design that (i) identifies a small subset of nodes in $G$ that are most relevant to the query, called \textit{attention nodes}, and subsequently (ii) computes
important statistics and performs graph traversal starting from attention nodes only. In particular, to ensure $\epsilon$-approximate result quality (defined in Section \ref{sec:problemdef}), it suffices to identify $O(\frac{1}{\eps})$ attention nodes. Existing solutions need to perform similar computations on a far larger set of nodes, covering the entire graph $G$ in the worst case.

Table~\ref{tbl:algos} compares the asymptotic performance of \ours against several recent approaches, 
where $n$ and $m$ denote the number of nodes and edges in $G$, respectively, and $\epsilon$ and $\delta$ are parameters for the error guarantee. For sparse graphs, $m$ is comparable to $O(n \log n)$; hence, compared to \probesim, the complexity of \ours is lower for common values of $\epsilon$ and $\delta$. Further, \ours does not involve large hidden constant factors (\textit{e.g.}, as in \sling), and makes no assumption on the data distribution of the underlying graph $G$ (\textit{e.g.}, as in \prsim, which assumes that $G$ is a power-law graph), as elaborated in Section \ref{sec:stateoftheart}.

\footnotetext[1]{ $O\left(n\log{\frac{n}{\delta}}/\epsilon^2\cdot\sum_{w\in V}{\pi(w)^2}\right)$ is the detailed time complexity of \prsim, where $\sum_{w\in V}{\pi(w)^2} = 1$ in the worst case \cite{prsimWeiHX0LDW19}.}

We experimentally evaluate our method against 6 recent solutions using \revise{9 real graphs}.
The results demonstrate the high practical performance of \ours. In particular, \ours outperforms all existing methods (both indexed and index-free) in terms of query processing time, and \ours is usually over an order of magnitude faster than the previous best index-free method {\sf ProbeSim}, on comparable result accuracy levels. \revise{Further, on UK graph with $133$ million nodes and $5.4$ billion edges, \ours obtains 0.00035 empirical error within $62$ milliseconds.}


\section{Preliminaries} \label{sec:prelim}

\subsection{Problem Definition} \label{sec:problemdef}

Let $G=(V, E)$ be a directed graph, where $V$ is the set of nodes with cardinality $n=|V|$, and $E$ is the set of edges with cardinality $m=|E|$. If the input graph is undirected, we simply convert each undirected edge $(u, v)$ to a pair of directed ones $(u, v)$ and $(v, u)$ with opposing directions. Following common practice in previous work
\cite{prsimWeiHX0LDW19,probsimLiuZHWXZL17}, we define the approximate single-source SimRank query as follows. Table \ref{tbl:notations} lists frequently used notations in the paper.

\begin{definition}\label{def:singlesourcequery}
(Approximate Single Source SimRank Query)
Given an input graph $G=(V, E)$, a query node $u \in V$, an absolute error threshold $\eps$, a failure probability $\pf$, and decay factor $c$, an approximate single source SimRank query returns an estimated value $\simap(u,v)$ for the exact SimRank $s(u, v)$ of each node $v \in V$, such that 
\begin{align}
|\simap(u,v) - \sim(u,v)| \leq \eps \label{eqn:singlesource}
\end{align}
holds for any $v\in V$ with at least $1-\pf$ probability.
\end{definition}



\subsection{State of the Art}\label{sec:stateoftheart}

\noindent
\textbf{{\sf \textbf{SLING}} \cite{slingTianX16}.} SimRank is well known to be linked to random walks \cite{jeh2002simrank}. Earlier work on SimRank processing generally use random walks without decay. More recent approaches are mostly based on a variant called $\sqrtc$-walks, as follows.

\begin{definition}
($\sqrtc$-Walk \cite{slingTianX16}) Given node $u$ and decay factor $c$, $\sqrtc$-walk from $u$ is a random walk that (i) has $1-\sqrtc$ probability to stop at current node, and (ii) has $\sqrtc$ probability to jump to a random in-neighbor of current node.
\end{definition}

Given two $\sqrtc$-walks from distinct nodes $u$ and $v$ respectively, we say that these two $\sqrtc$-walks \textit{meet}, if they both reach the same node after the same number of steps, say, the $\ell$-th step. 
Let $\lml(u,v,w)$ be the probability that two $\sqrtc$-walks from $u$ and $v$ meet at $w$ at the $\ell$-th step, and \textit{never meet again afterwards}.
Ref. \cite{slingTianX16} interprets  the SimRank value $\sim(u,v)$ as follows:
\begin{equation}\label{eq:simslingLM}
    \sim(u,v)=\sum_{\ell=0}^{+\infty}\sum_{w\in V} \lml(u,v,w).
\end{equation}

\sling~\cite{slingTianX16} further decomposes $\lml(u,v,w)$ into the product of three probabilities:
\begin{equation}\label{eq:simsling}
\lml(u,v,w) = \hl(u,w)\cdot\eta(w)\cdot\hl(v,w),
\end{equation}
where $\hl(u,w)$ denotes the probability (called \textit{hitting probability}) that a $\sqrtc$-walk from node $u$ reaches node $w$ at the $\ell$-th step. Since the random walks starting from nodes $u$ and $v$ are independent, the product $\hl(u,w)\cdot\hl(v,w)$ gives the probability (called \textit{meeting probability}) that these two walks meet at node $w$ (called the \textit{meeting node}). The correction factor $\eta(w)$ (called the \textit{last-meeting probability} of node $w$) is the probability that the above two $\sqrtc$-walks, after meeting at $w$, never meet again in the future. Clearly, this is equivalent to the probability that two independent $\sqrtc$-walks starting from $w$ never meet at any step.

\begin{table}[!t]
\centering
\begin{footnotesize}
\renewcommand{\arraystretch}{1.2}
\vspace{0mm}
\caption{Frequently used notations.} \label{tbl:notations}
\vspace{-3mm}
\begin{tabular}{|p{0.65in}|p{2.4in}|}
    \hline
    {\bf Notation} &  {\bf Description}\\
    \hline
    $G=(V,E)$   & Input graph $G$ with nodes $V$ and edges $E$\\
    \hline
    $n,m$   & $n=|V|, m=|E|$\\
    \hline
    $\outN(v),\inN(v)$ & Out-neighbors and in-neighbors of node $v$ \\
    \hline
    $\outD(v),\inD(v)$ & Out-degree and in-degree of node $v$ \\
    \hline
    $c$ & Decay factor in SimRank \\
    \hline
    $\eps$, $\delta$ & Maximum absolute error and failure probability in approximate SimRank\\
    \hline
    $\epsp$ & Error parameter decided by $\eps$ and $c$\\
    \hline
    $\Tsrc$ & Source graph generated for query node $u$ \\
    \hline
     $\Aset$ & Set of all attention nodes with respect to $u$ \\
     \hline
    $\Asetl$ & Set of attention nodes at the $\ell$-th level of  $\Tsrc$, where $\ell=1,...,L$ \\
    \hline
    $L$ & Max level in $\Tsrc$ \\
    \hline
    $\wl$, $\wli$, $\wlj$ & Attention nodes at the $\ell$-th level, $(\ell+i)$-th level, and $(\ell+j)$-th level of $\Tsrc$ respectively, where $\ell=1,...,L$ and $i=0,...,L-\ell$ \\
    \hline
    $\hl(u,w)$ & $\ell$-step hitting probability from $u$ to $w$ in $G$  \\
    \hline
    $\hlap(u,w)$ &  $\ell$-step hitting probability from $u$ to $w$ in $\Tsrc$  \\
    \hline
    $\hrlap(v,w)$ & Approximate hitting probability from $v$ to $w$ in $G$  \\
    \hline
     $\corrl(w)$& Last-meeting probability of attention node $w$ at the $l$-th level of $\Tsrc$\\   
      \hline
    $\resil(w)$ & Residue of attention node $w$, $\resil(w) = \hl(u,w) \cdot \corrl(w)$\\
    \hline
    \vspace{-2mm}
    \revise{$\lml(u,v,w)$} &
    \vspace{-2mm}\revise{The probability that two $\sqrtc$-walks from $u$ and $v$ meet at $w$ at the $\ell$-th step, and never meet again afterwards.}\\
    \hline
\end{tabular}
\vspace{-4.5mm}
\end{footnotesize}
\end{table}

\sling then pre-computes $\hl(u,w)$ and $\eta(w)$ with error up to $\eps$, and materializes them in its index. Given a query node $u$, \sling retrieves all nodes at all levels with $\hl(u,w)\geq \eps$. Then, for each level $\ell$ and every node $w$ on the $\ell$-th level, \sling 
retrieves $\eta(w)$ and each node $v$ with $\hl(v,w)\geq\eps$, and estimates $\sim(u,v)$ using Equation \eqref{eq:simsling}. 

\sling incurs substantial pre-processing costs for computing $\hl(u,w)$ and $\eta(w)$, which need to be re-computed whenever graph $G$ changes, as there is no clear way to efficiently update them. Consequently, \sling is not suitable for online processing. Further,  although \sling achieves beautiful asymptotic bounds as shown in Table \ref{tbl:algos}, its practical performance tends to be sub-par due to large hidden constant factors. For instance, Ref. \cite{prsimWeiHX0LDW19} points out that the index size of \sling is over an order of magnitude larger than $G$ itself, which leads to high retrieal costs at query time. Our experiments in Section \ref{sec:experiments} lead to similar conclusions.

\vspace{2mm}
\noindent 
{\sf \textbf{PRSim}}\textbf{\cite{prsimWeiHX0LDW19}.} \prsim is based on the main concepts of \sling, and further optimizes  performance, especially for \textit{power-law graphs}.  \prsim builds a connection between SimRank and personalized PageRank \cite{JehW03}: let $\pi^{(\ell)}(u,w)$ be the $\ell$-hop reverse personalized PageRank (RPPR) between  $u$ and $w$, we have $\pi^{(\ell)}(u,w) = \hl(u,w) \cdot (1-\sqrtc)$. \prsim uses Equation \eqref{eq:simprsim} for SimRank estimation:
\begin{equation}\label{eq:simprsim}
    \sim(u,v)=\frac{1}{(1-\sqrtc)^2}\sum_{\ell=0}^{+\infty}\sum_{w\in V} \pi^{(\ell)}(u,w)\cdot\eta(w)\cdot\pi^{(\ell)}(v,w). 
\end{equation}

Then, based on the assumption that the input graph $G$ is a power-law graph, \prsim selects a number of hub nodes, and pre-computes their RPPR values. At query time, \prsim estimates $\pi^{(\ell)}(u,w)\cdot\eta(w)$ by generating $\sqrtc$-walks from $u$ and $w$.
If $w$ happens to be a hub, \prsim seeks the index for all possible $\pi^{(\ell)}(v,w)$ for any $v\in V$; otherwise, $\pi^{(\ell)}(v,w)$ is estimated online using a sampling based technique.
Finally, \prsim estimates $\sim(u,v)$ based on Equation \eqref{eq:simprsim}.



Similar to \sling, \prsim incurs considerable pre-computation as explained above, 
and hence, it is not suitable for online SimRank processing. Further, \prsim heavily relies on the power-law graph assumption, both in algorithm design and in its asymptotic complexity analysis. In particular, in the best case that the underlying graph $G$ strictly follows power-law, the query time complexity is sublinear to the graph size \cite{prsimWeiHX0LDW19}. However, this assumption is rather strong and might be unrealistic: as reported in a recent study \cite{powerlawisrare}, strict power-law graphs are rare in practice.


\vspace{2mm}
\noindent 
{\sf \textbf{ProbeSim}}\textbf{ \cite{probsimLiuZHWXZL17}.} The state-of-the-art index-free method is \probesim. Specifically, 
let $W(u)$ and $W(v)$ be two $\sqrtc$-walks from nodes $u$ and $v$, respectively, and
 $f^{(\ell)}(u,v,w)$ be the probability that $W(u)$ and $W(v)$ \textit{first meet} at $w$ at the $\ell$-th step.
\probesim employs Equation \eqref{eq:simprobsim} to estimate SimRank:
\begin{equation}\label{eq:simprobsim}
      s(u,v) = \sum_{\ell=0}^{+\infty}\sum_{w\in V} f^{(\ell)}(u,v,w).
\end{equation}

Given query node $u$, \probesim first samples a $\sqrtc$-walk $W(u)$ from $u$. For every node $w$ 
at the $\ell$-th step of the walk, \probesim performs a probing procedure, in order to compute the first meeting probabilities at all levels. In particular, \probesim probes nodes in the order of increasing steps, so that when probing $w$ at the $\ell$-th step of $W(u)$, the method excludes the nodes visited in previous probings, in order to compute the first meeting probabilities in Equation \eqref{eq:simprobsim}. Such inefficiency leads to long query response time, which may put off users who wait online for query results.

\vspace{2mm}
\noindent 
\textbf{Other methods.}  \reads~\cite{readsJiangFWW17} precomputes $\sqrtc$-walks and compresses the walks into trees. During query processing, \reads retrieves the walks originating from the query node $u$, and finds all the $\sqrtc$-walks that meet with the $\sqrtc$-walks of $u$.
\tsf\cite{tsfShaoC0LX15} builds an index consisting of 
\textit{one-way graphs} by sampling one in-neighbor from each node's in-coming edges. During query processing, the one-way graphs are used to simlulate random walks to estimate SimRank. According to \cite{prsimWeiHX0LDW19}, \prsim subsumes both READS and TSF; further, \cite{prsimWeiHX0LDW19} points out that the result quality guarantee of \tsf is questionable, since (i) \tsf allows two walks to meet multiple times, leading to overestimated SimRank values and (ii) \tsf assumes that a random walk has no cycles, which may not hold in practice. Finally, 
\topsim\cite{topsimLeeLY12} is another index-free method, which is subsumed by \probesim according to \cite{probsimLiuZHWXZL17}. Meanwhile, according to \cite{probsimLiuZHWXZL17, prsimWeiHX0LDW19}, the result quality guaranee of \topsim is problematic as the method truncates random walks with a maximum number of steps.



\begin{figure*}[t!]
    \begin{subfigure}{0.65\textwidth}
        \centering
        \includegraphics[width=\textwidth]{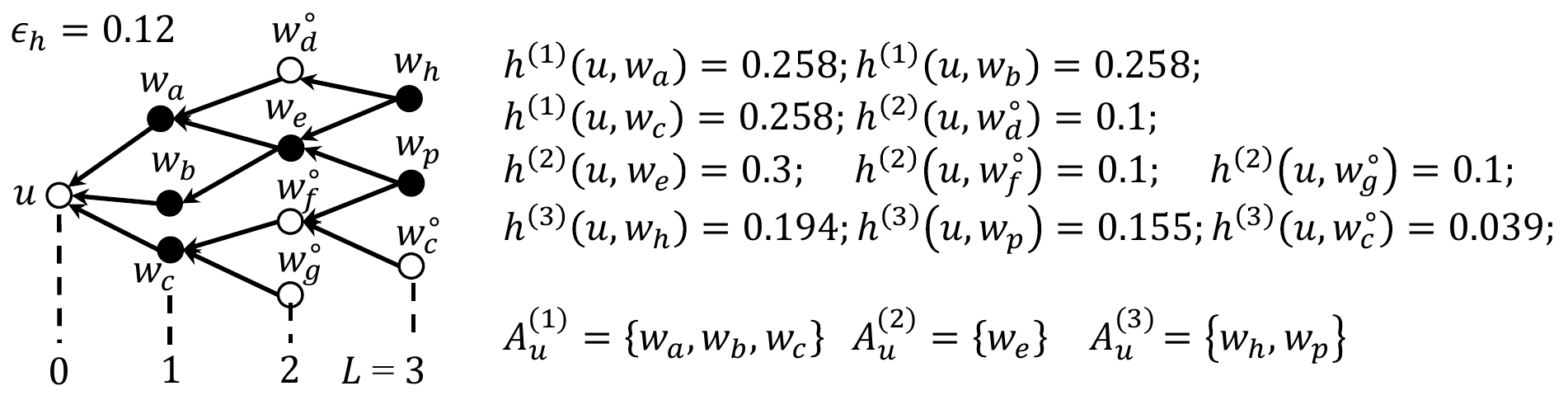}
        \vspace{-6mm}
        \caption{Source graph $\Tsrc$ and attention sets $\Asetl$: attention nodes are in black.} \label{fig:tsrcexample}
    \end{subfigure}%
    \hfill
    \begin{subfigure}{0.28\textwidth}
        \centering
        \includegraphics[width=0.95\textwidth]{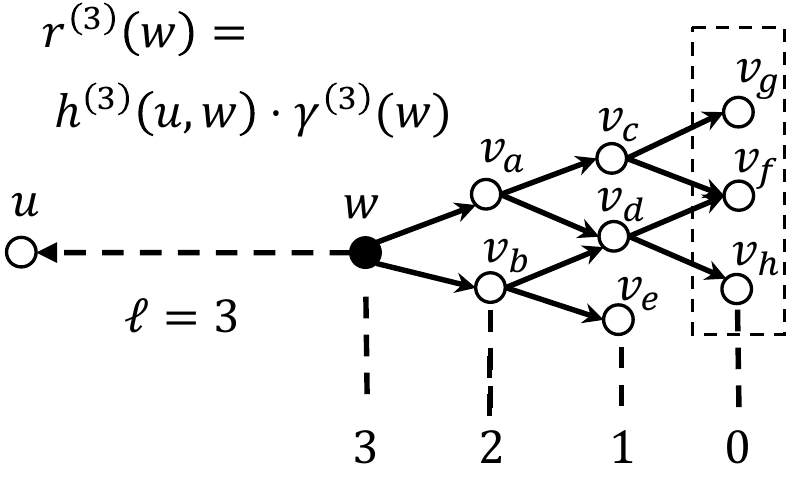}
        \vspace{-1mm}
        \caption{\revspread from $w$}
        \label{fig:revpushexample}
    \end{subfigure}
    \vspace{-2.5mm}
    \caption{Running Example of \ours}
    \label{fig:tsrcAndRevPushExamples}
        \vspace{-4mm}
\end{figure*}

\section{Overview of SimPush}\label{sec:overview}




We overview the proposed solution \ours  in this section, and present the detailed algorithm later in Section \ref{sec:mainalgo}. As mentioned before, the main idea of \ours is to identify a small set of \textit{attention nodes}, and focus computations  around these  nodes only. As we show soon, the number of attention nodes is bounded by $O(\frac{1}{\eps})$, and they are mostly within the close vicinity of the query node $u$, meaning that they can be efficiently identified. Meanwhile, we prove that the error introduced by neglecting non-attention nodes is negligible and bounded within the error guarantee $\epsilon$ in Inequality \eqref{eqn:singlesource}. This design significantly reduces the computational overhead in \ours.


Specifically, given the input graph $G$ and query node $u$, \ours computes the approximate single source SimRank results for $u$ in three stages. The first stage identifies the set of attention nodes, denoted as $A_u$, through a \textit{\leftpushnospace} algorithm. Besides $A_u$, \leftpushnospace also returns a graph $G_u$ (referred to as the \textit{source graph} of $u$) consisting of nodes in $G$ that are visited during the algorithm. In the second stage, \ours follows a similar (and yet much improved) framework as \sling, and computes the hitting probabilities between the query node $u$ and each attention node $w \in A_u$, as well as the last-meeting probability of $w$. \revise{Note that in \ours, the computation of hitting probabilities is restricted to attention nodes, and heavily reuses the intermediate results obtained in the first stage, which drastically reduces the computational overhead compared to existing methods such as \sling, which precomputes hitting probabilities for all nodes in a graph by following out-going edges. Further, \ours defines last-meeting probabilities over attention nodes only, and computes the probabilities in a deterministic way over a small \textit{source graph} generated when computing the attention nodes (details in Section 4.1), while previous methods such as \sling defines its last-meeting probabilities over the whole graph, and precomputes the probabilities by sampling numerous $\sqrtc$-walks.} Finally, in the third stage, \ours employs a \textit{\revspreadnospace} approach to complete the estimates of  probabilities between the query node $u$ and every node $v \in G$ via an attention node $w\in A_u$, yielding the final estimate of the SimRank between $u$ and $v$. In the following, we elaborate on  the three stages using the running example in Figure \ref{fig:tsrcAndRevPushExamples}.




\header
\textbf{Discovery of attention nodes.}
 First we clarify what qualifies a node as an attention node of query node $u$.

\vspace{-1mm}
\begin{definition}\label{def:attentionnode}
(Attention Nodes on Level $\ell$).
Given an input graph $G$ and a query node $u \in G$, a node $w$ is an attention node of $u$ on the $\ell$-th level, if and only if hitting probability 
$\hl(u,w)\geq\epsp,$
where $\epsp=\frac{1-\sqrtc}{3\sqrtc}\cdot \eps$.
\vspace{-1mm}
\end{definition}


Parameter $\epsp$ is explained in Lemma \ref{lemma:simoursEpsp} towards the end of this subsection. Let $\Asetl$ denote the set of attention nodes on level $\ell$, and $\Aset$ be the set of all attention nodes that appear in any level. 
Focusing on the attention nodes only, we employ the interpretation of SimRank $s(u, v)$ in Equation \eqref{eq:simslingLM}, and have the approximate $s'(u, v)$ in Equation \eqref{eq:simoursS1Aset}.
Lemma \ref{lemma:simoursS1Aset} provides the error guarantee for  $s'(u, v)$ \footnote{All proofs can be found in the appendix.}.


\begin{equation}\label{eq:simoursS1Aset}
    \simapleft(u,v) =\sum_{\ell=0}^{+\infty}\sum_{w\in \Asetl} \lml(u,v,w), 
\end{equation}

\begin{lemma}\label{lemma:simoursS1Aset}
Given nodes $u, v \in G$, their exact SimRank $s(u,v)$ and estimated value $\simapleft(u,v)$ in Equation \eqref{eq:simoursS1Aset} satisfy
$$s(u,v)-\frac{\sqrtc\cdot\epsp}{1-\sqrtc}\leq \simapleft(u,v) \leq s(u,v).$$
\end{lemma}

In the above definition of $s'(u, v)$, we enumerate all possible levels $\ell$. Next we show that this is not necessary, since attention nodes only exist in the first few levels within close vicinity of query node $u$, according to the following lemma.

\begin{lemma}\label{lemma:AsetandL}
Given query node $u$, decay factor $c$ and parameter $\epsp$, the number of attention nodes with respect to $u$ is at most $\left\lfloor\frac{\sqrtc}{(1-\sqrtc)\cdot\epsp}\right\rfloor$. Meanwhile, all attention nodes exist within $L^*=\left\lfloor\log_{\frac{1}{\sqrtc}}\frac{1}{\epsp}\right\rfloor$ steps from $u$.
\end{lemma}


According to Lemma \ref{lemma:AsetandL}, to discover all attention nodes, it suffices to explore $L^*$ steps around the query node $u$. Further, in \ours, attention node discovery is performed by exploring $L\leq L^*$ steps from $u$, through the proposed \leftpush algorithm, detailed in Section \ref{sec:leftpush}.
In particular, \leftpush samples a sufficient number of random walks to determine $L$, such that with high probability (according to parameter $\delta$), all attention nodes exist within $L$ steps from $u$. The specific value of $L$ depends on the input graph $G$. As our experiments demonstrate $L$ is usually small for real graphs. For instance, when $\epsilon=0.02$, on a billion-edge Twitter graph, the average $L$ is merely $2.76$, and the number of attention nodes is no more than a few hundred. 

Next, to identify attention nodes, \ours also needs to compute the hitting probabilities from $u$. \revise{This is done through a {\it residue propagation} procedure in the  \leftpush algorithm, detailed in Section \ref{sec:leftpush}. Specifically, $h^{(0)}(u,u)$ is set to 1, and all other hitting probabilities are initialized to zero. Starting from the $0$-th level, \leftpush pushes hitting probabilities of nodes from the current level to their in-neighbors on the next level, until reaching the $L$-th level.}
As mentioned earlier, \ours also records the nodes and edges traversed during the propagation in a \textit{source graph} $\Tsrc$. Specifically, 
$\Tsrc$ is organized by levels (with max level $L$), and there are only edges between adjacent levels, \textit{i.e.}, incoming edges from the $(\ell+1)$-th level to the $\ell$-th level. 
$\Tsrc$ itself, as well as the computed hitting probabilities of attention nodes, are reused in subsequent stages of \ours.

Figure \ref{fig:tsrcexample} shows an example of the propagation process, assuming $L=3$ and $\epsp=0.12$. Attention (resp.\ non-attention) nodes are shown as solid circles (resp.\ empty circles) in the figure. \revise{Symbols with a superscript circle ({\it e.g.}, $w_d^{\circ}$) denote non-attention nodes, which are used later in Section \ref{sec:mainalgo}}.
Specifically, the propagation starts from $u$ and traverses the graph in a level-wise manner, reaching nodes $w_a, w_b, w_c$ on the first level, nodes $w_d, w_e, w_f, w_g$ on the second level, and nodes $w_h, w_p, w_c$ on the third level, which is the last level since $L=3$. Note that a node can be visited multiple times on different levels, {\it e.g.}, $w_c$ on both the first and third levels. In this case, it is also possible that a node is an attention node on one level (\textit{e.g.}, $w_c$ on Level 1) and non-attention node on another (\textit{e.g.}, $w_c$ on Level 3).


\header
\textbf{Estimation of $\lml(u,v,w)$.}
After identifying  attention nodes,  \ours needs to estimate each $\lml(u,v,w)$, according to Equation \eqref{eq:simoursS1Aset}. Existing solutions mostly estimate it by running numerous $\sqrtc$-walks on the whole graph $G$, which is costly. Instead, \ours incorporates a novel algorithm that mostly operates within the source graph $\Tsrc$ obtained in the first phase. $\Tsrc$ is far smaller than $G$.

Specifically, the hitting probabilities from $u$ to all attention nodes are already obtained Phase 1. Next, we focus on the  \textit{last meeting probability} for a given node $w$. In order to achieve high efficiency, \ours only computes last meeting probabilities for attention nodes, and limits the computations within the source graph $G_u$. Towards this end, \ours defines a new last meeting probability, as follows.

\begin{definition}\label{def:nevermeetprobintree}
(Last-Meeting Probability in $\Tsrc$).
Given attention node $\wl$ on the $\ell$-th level of $\Tsrc$, where $\ell=1, \ldots, L$, the last-meeting probability of $\wl$ within $\Tsrc$, $\corrl(w)$, is the probability that two $\sqrtc$-walks from $\wl$ and walking within $\Tsrc$ do not meet at any \textit{attention node} on the $(\ell+i)$-th level within $\Tsrc$, for  $1\leq i\leq L-\ell$.
\end{definition}


We emphasize that $\corrl(w)$ has vital differences compared to the last-meeting probability $\eta(w)$ used in \sling and \prsim, explained in Section \ref{sec:stateoftheart}.
First, $\corrl(w)$ is defined based on the attention sets and source graph $\Tsrc$, instead of the whole graph.
Second, $\corrl(w)$ does not take into account whether or not two walks meet at any non-attention node; the rationale here is that non-attention nodes have negligible impact on the SimRank estimation of \ours, and, thus, can be safely ignored.
Third, $\corrl(w)$ is \textit{level-specific} and we only consider $L-\ell$ steps in $\Tsrc$ since there are only incoming edges between consecutive levels in $\Tsrc$ and the levels are bounded by $L$. In Section \ref{sec:nevermeetcorr}, we present an efficient residue-push technique to compute the $\corrl(w)$ of all attention nodes, without performing any $\sqrtc$-walk.

Based on the above notion of last meeting probability, we design another estimate for the SimRank value $s(u, v)$ between the query node $u$ and a node $v \in G$, as follows.

\begin{equation}\label{eq:simoursS1gamma}
    \simapleftgamma(u,v) =\sum_{\ell=1}^{L^*}\sum_{w\in \Asetl} \hl(u,w)\cdot\corrl(w)\cdot\hl(v,w), 
\end{equation}
where $\Asetl$ is the set of attention nodes at the $\ell$-th level of   $G_u$, obtained in the first phase. Note that here the trivial case of $\ell=0$ is not considered, and we require $u \neq v$.

Compared to $s'(u, v)$ defined in Equation \eqref{eq:simoursS1Aset}, $s^+(u, v)$ uses an estimated $\lml(u,v,w)$, computed using hitting probabilities and last-meeting probabilities in $G_u$. The following lemma establishes the approximation bound for $s^+(u, v)$.


\begin{lemma}\label{lemma:simoursS1gamma}
Given distinct nodes $u$ and $v$, their exact SimRank value $s(u,v)$ and estimate $\simapleftgamma(u,v)$ satisfy
$$s(u,v)-\frac{2\sqrt{c}\cdot\eps_h}{1-\sqrt{c}}\leq \simapleftgamma(u,v)\leq s(u,v).$$
\end{lemma}

\header
\textbf{\revspreadnospace}.
In Equation \eqref{eq:simoursS1gamma}, it remains to clarify the computation of $\hl(v,w)$.
Instead of estimating $\hl(v,w)$ independently (\textit{e.g.}, by simulating random walks), we propose a novel \revspread algorithm, detailed in Section \ref{sec:revspread}, which estimates $\hl(u,w)\cdot\corrl(w)\cdot\hl(v,w)$ as a whole through residue push.
Specifically, \ours regards $\resil(w)=\hl(u,w)\cdot \corrl(w)$ as the initial residue of attention node $w$, and keeps pushing the residue to each node $v \in G$, following out-going edges, until $\ell$ steps are performed.

For example, in Figure \ref{fig:revpushexample}, given a $3$rd level attention node $w$ with residue $\resi^{(3)}(w)$, \revspread propagates the residue to the out-neighbors of $w$, \textit{i.e.}, $v_a$ and $v_b$, to obtain the residues at the $2$nd level, \textit{i.e.}, $\resi^{(2)}(v_a)$ and $\resi^{(2)}(v_b)$. Then, all $\resi^{(2)}$ residues are pushed to their out-neighbors to get all $\resi^{(1)}$ residues. After that, all $\resi^{(1)}$ are pushed to get $\resi^{(0)}$ residues.
It is clear that the nodes at the $0$-th level, \textit{e.g.}, $v_g$ (as well as $v_h$ and $v_k$) meets with $u$ at $w$ in 3 steps. The residue $\resi^{(0)}(v_g)$ estimates $h^{(3)}(u,w)\cdot\corr^{(3)}(w)\cdot h^{(3)}(v_g,w)$ w.r.t., $\resi^{(3)}(w)$.
The detailed push criteria is in Section \ref{sec:revspread}. 

\begin{table}[!t]
\centering
\begin{footnotesize}
\renewcommand{\arraystretch}{1.2}
\caption{Complexity of different stages in \ours.} \label{tbl:complexbystages}
\vspace{-3mm}
\begin{tabular}{|p{1.4in}|p{1.5in}|}
    \hline
    {\bf Stage} &  {\bf Time Complexity}\\
    \hline
    \leftpush   & $O(\complexleftpush)$\\
    \hline
    All $\corrl(w)$ computation  & $O(\complexlastmeetall)$\\
    \hline
    \revspread  & $O(\complexrevpush)$\\
    \hline
\end{tabular}
\vspace{-2mm}
\end{footnotesize}
\end{table}

Accordingly, our final SimRank estimate is

\begin{equation}\label{eq:simours}
    \simap(u,v) =\sum_{\ell=1}^{L^*}\sum_{w\in \Asetl} \hl(u,w)\cdot\corrl(w)\cdot\hrlap(v,w), 
\end{equation}
where $u$ and $v$ are distinct nodes in $G$, $\Asetl$ is the $\ell$-th level attention set. Here, the hitting probability $\hrlap(v,w)$ from $v$ to $w$ is hatted to signify that \revspread introduces additional estimation error. Note that as described above, the estimation is over the entire product $\hl(u,w)\cdot\corrl(w)\cdot\hl(v,w)$ rather than the last term.
Lemma \ref{lemma:simoursEpsp} provides error guarantee for $\simap(u,v)$, and explains the value of $\epsp$.

\begin{lemma}\label{lemma:simoursEpsp}
Given distinct nodes $u$ and $v$ in $G$, error parameter $\eps$, and decay factor $c$, when $\eps_h\leq \frac{1-\sqrt{c}}{3\sqrt{c}} \cdot \eps$, we have $s(u,v)-\simap(u,v) \leq \epsilon$.
\end{lemma}

Note that in Lemma \ref{lemma:simoursEpsp}, the error bound is deterministic, rather than probabilistic as in our problem definition in Inequality \eqref{eqn:singlesource}. This is due to the fact that in Equation \eqref{eq:simours}, we enumerate up to $L^*$ levels instead of $L$ levels as in the actual algorithm, as mentioned earlier. The value of $L$, as well as the probabilistic error bound of the complete \ours solution, are deferred to the next section. Finally, Table \ref{tbl:complexbystages} lists the  time complexity of the three stages of \ours.

\section{Detailed SimPush Algorithm}\label{sec:mainalgo}

\begin{algorithm}[!t]
\begin{small}
\caption{\ours} \label{algo:main}
\BlankLine
\KwIn{Graph $G=(V, E)$, query node $\src$, decay factor $\c$, error parameter $\eps$, failure probability $\delta$}
\KwOut{$\simap(u,v)$ for each $v\in V$, w.r.t, query node $u$.}
$\epsp\gets \frac{1-\sqrtc}{3\sqrtc}\cdot \eps$\;
Invoke Algorithm \ref{alg:leftpush} (\leftpushnospace) to obtain attention nodes and the source graph $\Tsrc$\;
Invoke Algorithm \ref{alg:pushintree} to compute all nonzero hitting probabilities for attention nodes in $\Tsrc$\;
\For{$\ell=1$ to $L$}{
    \For{each attention node $w$ in $\Asetl$}{
    Compute $\corrl(w)$ with Algorithm \ref{alg:nevermeetcorr}\;
    $\resil(w)\gets \hl(u,w)\cdot\corrl(w)$\;
    }
}

Invoke Algorithm \ref{alg:revspread} (\revspreadnospace) to get $\simap(u,v)$ for each $v\in V$\;
\Return $\simap(u,v)$ for each $v\in V$\;
\end{small}
\end{algorithm}

Algorithm \ref{algo:main} shows the main \ours algorithm, consisting of three stages. With $\epsp$ set at Line 1, \ours first invokes \leftpush (Section \ref{sec:leftpush}) to obtain the attention nodes  and source graph $\Tsrc$ of $u$  (Line 2). Then (Lines 3-7), it computes the $\corrl(w)$ of all attention nodes $w$ (Section \ref{sec:nevermeetcorr}), and finally invokes \revspread (Section \ref{sec:revspread})  to compute the single source SimRank values at Line 8.

\subsection{\leftpush}\label{sec:leftpush}


\leftpush first samples a sufficient number of random walks to detect the max level $L$ from query node $u$, such that with high probability, all attention nodes appear within $L$ steps. Then, it performs residue propagation to compute the hitting probabilities from $u$, in order to identify attention nodes of $u$ and generate source graph $\Tsrc$.
Algorithm \ref{alg:leftpush} displays  \leftpushnospace.
At Lines 1-3, \leftpush first samples $\left(\numWalks\right)$ $\sqrtc$-walks from $u$, counts the  visits of every node $v$ at every $l$-th step,  $H^{(l)}(u,v)$, and then identifies the max level $L$ where there exists node $v$ with $H^{(l)}(u,v)\geq\left(\numwalkshalf\right)$, and $L$ is bounded by $L^*$ (Lines 4-8).
Then, Algorithm \ref{alg:leftpush} computes the hitting probabilities from $u$ for at most $L$ levels by propagation (Lines 9-19).
Initially, at Lines 9-10, $\hzero(u,u)$ is set to 1, all other hitting probabilities are initialized to zero.
Starting from the $0$-th level,  \leftpush inserts $u$ into frontier set $F$ at Line 11, and then for each node $v$ in $F$ at the current $\ell$-th level, it pushes and increases the $(\ell+1)$-level hitting probability of every in-neighbor $v^\prime$ of $v$ by $\frac{\sqrtc\cdot\hl(u,v)}{\inD(v)}$ and adds edge from $v^\prime$ to $v$ to $\Tsrc$ (Lines 12-16).
Then, \leftpush moves to the $(\ell+1)$-th level, and finds all the nodes to push (Lines 17-19). The whole process continues until the $L$-th level is reached or $F$ is empty (Line 12).
At Lines 20-21, all attention nodes are identified.
Lemma \ref{lemma:leftpush} states the accuracy guarantees and time complexity of Algorithm \ref{alg:leftpush}.


\begin{lemma} \label{lemma:leftpush}
Algorithm \ref{alg:leftpush} runs in $O(\complexleftpush)$ expected time, and with probability at least $1-\delta$, $\Tsrc$ contains all nodes $w$ with $\hl(u,w)\geq \epsp$ for all levels.
\end{lemma}


\begin{algorithm}[!t]
\begin{small}
\caption{$\mathsf{\leftpush}$}\label{alg:leftpush}
\BlankLine
\KwIn{Graph $G$, query $\src$, decay factor $c$, parameter $\epsp$}
\KwOut{Source graph $\Tsrc$ and attention node sets $\Asetl$ for $\ell=1,...,L$.}

$H^{(l)}(u,v)\gets 0$, for $v\in V$ and $l=1,2,...$\;
\For{$i=1,...,\left(\numWalks\right)$}{
    Generate a $\sqrtc$-walk from $u$ and for every visited node $v$ at the $l$-th step, $H^{(l)}(u,v)\gets H^{(l)}(u,v)+1$\;
}
$L\gets 0$\;
\For{every nonzero $H^{(l)}(u,v)$}{
    \If{$l>L$\text{ and }$H^{(l)}(u,v)\geq \numwalkshalf $}{
        $L\gets l$\;
    }
}
$L\gets \min(L,L^*)$\;
$\hl(\src, v)\gets 0$ for $\ell=1,...,L$ and each $v\in V$\; 
$\ell\gets 0$; $\hzero(\src, \src) \gets 1$\; 
Frontier set $F\gets\{\src\}$\;
\While {$F\neq \emptyset$ \text{ and } $\ell< L$}{
    \For{each $v \in F$}{
        \For {each node $v^\prime \in \inN(v)$}{
	        $\hlnext(\src,v^\prime) \gets \hlnext(\src,v^\prime) + \frac{\sqrtc\cdot\hl(\src,v)}{\inD(v)}$\;
	        Insert $v$ to the $\ell$-th level and $v'$ to the $(\ell+1)$-th level of $\Tsrc$, and add edge from $v'$ to $v$ in $\Tsrc$\;
	    }
    }
    $F\gets \emptyset$;
    $\ell \gets \ell + 1$\;
    \For{each node $v$ with $\hl(\src,v)>0$} {
        $F\gets F\cup\{v\}$\;
    } 
}
\For{$\ell = 1,...,L$}{
    Insert $w$ in $\Tsrc$ with $\hl(u,w)\geq \epsp$ into $\Asetl$\;
}
\end{small}
\end{algorithm}

Lastly, we define hitting probability within $\Tsrc$, which is an important concept used in the next stages of \ours.

\begin{definition}\label{def:hpintree}
(Hitting probability in  $\Tsrc$).
Given nodes $w_a$ and $w_b$ in $\Tsrc$, the hitting probability from $w_a$ to $w_b$  at the $i$-th step in $\Tsrc$, is the probability that a $\sqrtc$-walk from $w_a$ and walking in $\Tsrc$, visits $w_b$ at the $i$-th step, where $i\geq 0$. 
\end{definition}

Hereafter, we use $\hlap(*,*)$ to denote the  hitting probabilities in $\Tsrc$, and use  $\hl(*,*)$ to represent the  hitting probabilities in $G$.
For query node $u$, every $\hl(u,w)$ computed by \leftpush over $G$ can be reproduced by pushing $u$ over $\Tsrc$, \textit{i.e.}, $\hlap(u,w)$ is the same as $\hl(u,w)$.
For the ease of presentation, in the following sections, we denote $\wl$, $\wli$, and $\wlj$ as  nodes at the $\ell$-th, $(\ell + i)$-th, $(\ell + j)$-th levels of $\Tsrc$ respectively, and  $\wl,\wli,\wlj$ are attention nodes by default, unless otherwise specified.

\subsection{\nevermeetcorr within $\Tsrc$} \label{sec:nevermeetcorr}

As mentioned, given query  $u$ with attention sets $\Asetl$, \ours computes last-meeting probability $\corrl(w)$ for each $w\in\Asetl$ in the source graph $\Tsrc$ (Definition \ref{def:nevermeetprobintree}).
Utilizing  $\Tsrc$, we design a method that computes $\corrl(w)$ for \textit{all} attention nodes in $\Tsrc$ without generating any $\sqrtc$-walks, in $O(\complexlastmeetall)$ time.
We first clarify the formula to compute $\corrl(w)$, and then present the detailed algorithms.


\header
\textbf{Formula to compute $\corrl(w)$.}
Given attention nodes $w$ and $\wli$, we define the $i$-step first-meeting probability $\meeti(w,\wli)$ in $\Tsrc$ as follows.
\begin{definition} \label{def:firstmeetingprob}
(First-meeting probability in $\Tsrc$).
Given attention nodes $\wl$ and $\wli$ at the $\ell$-th and $(\ell+i)$-th levels of $\Tsrc$ respectively, where $\ell = 1,...,L$ and $0<i\leq L-\ell$,
$\meeti(w,\wli)$ is the probability that
two $\sqrtc$-walks from $\wl$ walking in $\Tsrc$ first meet at attention node  $\wli$.
\end{definition}

Note that in Definition \ref{def:firstmeetingprob}, it is allowed that the two walks first meet at some non-attention node in $\Tsrc$, before meeting at $\wli$.
In this section, when we say that two walks first meet, it means that the two walks first meet at an attention node in $\Tsrc$. 
According to Definitions \ref{def:nevermeetprobintree} and \ref{def:firstmeetingprob},  we have

\begin{equation}
\label{eqdefcor}
    \corrl(w) = 1- \sum_{i=1}^{L-\ell} \sum_{\wli\in \Asetli} \meeti(w,\wli),
\end{equation}
where $\Asetli$ is the $(\ell+i)$-th level attention set and $\ell\leq L$.
Now, the problem reduces to computing $\meeti(w,\wli)$ in $\Tsrc$. This requires the hitting probabilities  between attention nodes within $\Tsrc$ (Definition \ref{def:hpintree}), to be clarified soon.

\begin{figure}[!t]
  \centering
  \includegraphics[width=0.95\columnwidth]{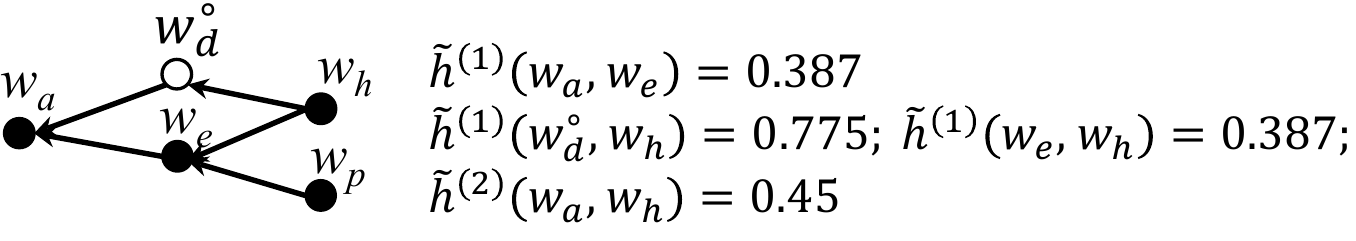}
  \vspace{-3mm}
  \caption{Hitting probabilities in a subgraph of $\Tsrc$ in Figure \ref{fig:tsrcexample}.}\label{fig:pushintree}
    \vspace{-2mm}
\end{figure}

When $i=1$, $\meetone(w,\wl_1)$ is nonzero only if attention node $\wl_1$ is an in-neighbor of $\wl$ in $\Tsrc$ (obviously $\wl_1$ is at the $(\ell+1)$-th level of $\Tsrc$).
Given the 1-step hitting probability $\hinT^{(1)}(w,\wl_1)$, the probability of two independent $\sqrtc$-walks from $\wl$ walking in $\Tsrc$ and meeting at $\wl_1$ is $\hinT^{(1)}(w,\wl_1)^2$. 
Further, since there is only one step from $\wl$ to $\wl_1$, $\meetone(w,\wl_1)$ is exactly $\hinT^{(1)}(w,\wl_1)^2$, \textit{i.e.}, 

\begin{equation}\label{eq:meetionelevel}
  \forall \wl_1\in \Aset^{(\ell+1)}, \meet^{(1)}(w,\wl_1) =\hinT^{(1)}(w,\wl_1)^2,
\end{equation}
where $\Aset^{(\ell+1)}$ is the $(\ell+1)$-th level attention set. For example, in Figure \ref{fig:pushintree}, $\meetone(w_a, w_e)=\hinT^{(1)}(w_a,v_e)^2=0.150$.

\begin{figure}[!t]
  \centering
  \includegraphics[width=0.45\columnwidth]{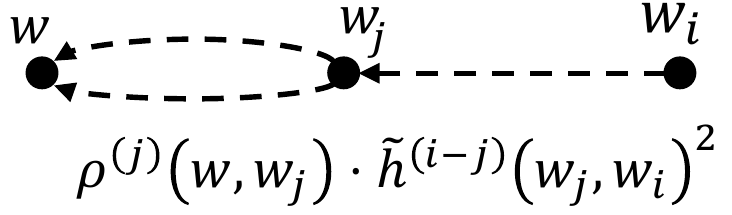}
  \vspace{-2mm}
  \caption{Non-first-meeting probability from attention nodes $\wl$ to $\wli$ via $\wlj$.}\label{fig:nonfirstmeeting}
  \vspace{-3mm}
\end{figure}

When $2\leq i\leq L-\ell$,
we compute $\meeti(w,\wli)$ by utilizing  $\meetj(w,\wlj)$ of the attention nodes $\wlj$ between $w$ and $\wli$ in $\Tsrc$, where  $1\leq j \leq i-1$.
Suppose that two $\sqrtc$-walks from $w$ walking in $\Tsrc$ first meet at $\wlj$ and then meet at $\wli$.
The \textit{non-first-meeting probability} from $w$ to $w_i$ via $w_j$ is 
$\meetj(w,\wlj)\cdot\hinT^{(i-j)}(\wlj,\wli)^2$. Figure \ref{fig:nonfirstmeeting} illustrates this concept, where first-meeting probability $\meetj(w,\wlj)$ is represented by two dashed lines, and meeting probability $\hinT^{(i-j)}(\wlj,\wli)^2$ is represented by one dashed line. Therefore, $\meeti(w,\wli)$ equals the meeting probability from $\wl$ to $\wli$, \textit{i.e.}, $\hinTi(w,\wli)^2$, subtracted by all the non-first-meeting probabilities from $\wl$ to $\wli$ via any attention node $\wlj$ between $w$ and $\wli$, \textit{i.e.}, 

\begin{equation}\label{eq:nonfirstmeetleveli}
\begin{split}
 \meeti(w,\wli) &= \hinTi(w,\wli)^2 \\ 
  - &\sum_{j=1}^{i-1}\sum_{\wlj\in\Asetlj} \meet^{(j)}(w,\wlj)\cdot \hinT^{(i-j)}(\wlj,\wli)^2,\\
\end{split}
\end{equation}
where $i=2,..., L-\ell$. For example, in Figure \ref{fig:pushintree},
$\meet^{(2)}(w_a,w_h)=h^{(2)}(w_a,w_h)^2-\meetone(w_a,w_e)\cdot h^{(1)}(w_e,w_h)^2=0.45^2 -0.15\cdot 0.387^2=0.18.$ $w_d^\circ$ is not considered since it is a non-attention node.

\header
\textbf{Hitting probabilities between attention nodes in  $\Tsrc$.}
Now we focus on computing hitting probabilities in $\Tsrc$.
Given nodes $\wl$ and $\wli$ (here $w$ can be a non-attention node), $\hinT^{(i)}(\wl,\wli)$ is computed by aggregating the hitting probabilities $\hinT^{(i-1)}(w',\wli)$ from $w$'s in-neighbors $w'$ to $\wli$, as follows.

\begin{equation}\label{eq:hinTcomp}
\hinT^{(i)}(\wl,\wli) =
\frac{\sqrtc}{\inDTu(\wl)}\sum\limits_{w'\in\inNTu(\wl)}{\hinT^{(i-1)}(w',\wli)},
\end{equation}
where $\inNTu(\wl)$ is the set of in-neighbors of $w$ in $\Tsrc$ and $\inDTu(\wl)$ is the indegree of $w$ in $\Tsrc$, and $i\geq 1$. For example, in Figure \ref{fig:pushintree}, $\hinT^{(2)}(w_a,w_h)=\frac{\sqrtc}{2}\cdot\left( \hinT^{(1)}(w^\circ_d,w_h) + \hinT^{(1)}(w_e,w_h)\right)=0.45$.
Note that (i) in Equation \eqref{eq:hinTcomp}, $w'$ can be a non-attention node if it has nonzero hitting probabilities to attention nodes in $\Tsrc$, \textit{e.g.}, $\hinT^{(1)}(w^\circ_d,w_h)$ in the example; (ii) if node $w$ has nonempty $\inNTu(\wl)$ in $\Tsrc$, $\inNTu(\wl)$ is the same as $\inN(\wl)$ in $G$.



\begin{algorithm}[!t]
\begin{small}
\caption{Hitting probabilities in $\Tsrc$}\label{alg:pushintree}
\BlankLine
\KwIn{Source graph $\Tsrc$}
\KwOut{All nonzero hitting probabilities between attention nodes in $\Tsrc$}

\For{$\ell=L,...,2$}{
    \For{each attention node $w$ at the $\ell$-th level of $\Tsrc$}{
       $\hzeroap(w,w)\gets 1$\;
    }
    \For{each node $w'$ at the $\ell$-th level of $\Tsrc$}{
        \For{each nonzero $\hinTi(w',\wli)$}{
            \For{each $w'_a \in \outNTu(w')$ at  $(\ell-1)$-th level}{
                $\hinT^{(i+1)}(w'_a,w_i)\gets \hinT^{(i+1)}(w'_a,w_i) + \frac{\sqrtc}{\inD(w'_a)}\cdot \hinTi(w',w_i)$
            }
        }
    }
}
\Return 
\end{small}
\end{algorithm}

\header
\textbf{Algorithms.}
Next we present two algorithms: Algorithm \ref{alg:pushintree} that computes the hitting probabilities between attention nodes within $\Tsrc$ using Equation \eqref{eq:hinTcomp}, and
Algorithm \ref{alg:nevermeetcorr} that computes $\corrl(w)$ using Equations \eqref{eqdefcor}, \eqref{eq:meetionelevel}, and \eqref{eq:nonfirstmeetleveli}.

In Algorithm \ref{alg:pushintree}, all hitting probabilities are initialized to zero.
Starting from $\ell=L$ to $2$, for each attention node $w$, we first set $\hzeroap(w,w)$ to 1 at Lines 2-3 (\textit{i.e.}, the hitting probability to itself is 1).
Then from Lines 4 to 7, for every node $w'$ at the $\ell$-th level (including non-attention nodes), if it has nonzero hitting probabilities  $\hinTi(w',w_i)$ to any attention node $\wli$ at the $(\ell+i)$-th level for $i=0,...,L-\ell$, each of the probabilities $\hinTi(w',w_i)$ is aggregated to every out-neighbor $w'_a$ of $w'$ in $\Tsrc$, where $w'_a$ is at the $(\ell-1)$-th level of $\Tsrc$ and can be a non-attention node. Apparently, from the perspective of $w'_a$, we are aggregating its in-neighbors' hitting probabilities to itself, \textit{i.e.}, Equation \eqref{eq:hinTcomp}.
Finally, only the hitting probabilities between attention nodes in $\Tsrc$ are returned and used by Algorithm \ref{alg:nevermeetcorr} for computing $\corrl(w)$.

Algorithm \ref{alg:nevermeetcorr} computes  $\corrl(w)$ for attention node $w$ at the $\ell$-th level of $\Tsrc$.
At Line 1, $\corrl(w)$ is initialized to 1. At Lines 2-4, when $i=1$, all nonzero $\meet^{(1)}(\wl,\wl_1)$ are computed according to Equation \eqref{eq:meetionelevel}, and are subtracted from $\corrl(w)$, based on Equation \eqref{eqdefcor}.
Then all first-meeting probabilities $\meeti$ for $2\leq i \leq \deltaL$ are computed level by level from Lines 5 to 11, using Equation \eqref{eq:nonfirstmeetleveli}.
Specifically, every $\meeti(w,\wli)$ is initialized as $\hinTi(w,\wli)^2$ at Lines 6-7 and is subtracted by all non-first meeting probabilities from $w$ to $\wli$ via $\wlj$ at Lines 8-11. 
Whenever the first-meeting probabilities $\meeti(\wl, \wli)$ for attention nodes $\wli\in \Asetli$ are obtained, they are subtracted from $\corrl(w)$ at Line 12, according to Equation \eqref{eqdefcor}.
Finally, $\corrl(w)$ is returned.
Lemma \ref{lemma:lastmeetingcomplex} presents the time complexity of Algorithm \ref{alg:pushintree}, Algorithm \ref{alg:nevermeetcorr}, and the second stage of \ours as a whole.

\begin{lemma}\label{lemma:lastmeetingcomplex}
Algorithm \ref{alg:pushintree} runs in $O(\complexpushintree)$ time, and Algorithm \ref{alg:nevermeetcorr} runs in $O(\complexlastmeetsingle)$ for a single $\corrl(w)$, and there are $O(1/\eps)$ attention nodes. Therefore, the overall time complexity for last-meeting  computation is $O(\complexlastmeetall)$.
\end{lemma}

\begin{algorithm}[!t]
\begin{small}
\caption{Last-Meeting Probability}\label{alg:nevermeetcorr}
\BlankLine
\KwIn{Source graph $\Tsrc$; attention node $w\in \Asetl$;}
\KwOut{Last-meeting probability $\corrl(w)$ in $\Tsrc$}
$\corrl(w)\gets 1$; $\deltaL\gets L-\ell$\;
\For{each attention node $w_1$ with nonzero $\hinT^{(1)}(w,w_1)$}{
    $\meet^{(1)}(w,w_1)\gets\hinT^{(1)}(w,w_1)^2$\;
}
$\corrl(w)\gets\corrl(w)-\sum_{w_1\in \Aset^{(\ell+1)}}\meet^{(1)}(w,w_1)$\;

\For{$i=2$ to $\deltaL$}{
    \For{each attention node $\wli$ with $\hinTi(w,\wli)>0$}{
        $\meeti(w,\wli)\gets\hinTi(w,\wli)^2$\;
    }
    \For{$j=1$ to $i-1$} {
        \For{each nonzero $\meet^{(j)}(w,\wlj)$ of each attention node $\wlj$ at the $(\ell+j)$-th level of $\Tsrc$}{
            \For{each nonzero $\hinT^{(i-j)}(\wlj,\wli)$}{
              $\meeti(w,\wli) \gets \meeti(w,\wli) - \meet^{(j)}(w,\wlj)\cdot \hinT^{(i-j)}(\wlj,\wli)^2$\;
            }
        }
    }
    $\corrl(w)\gets \corrl(w) - \sum_{\wli\in \Asetli}\meeti(w,\wli)$\;
}
\Return $\corrl(w)$\;
\end{small}
\end{algorithm}


\vspace{-2mm}
\subsection{\revspread}\label{sec:revspread}

Given $w$ in $\Asetl$ with its $\corrl(w)$ obtained, we regard $\resil(w)=\hl(u,w)\cdot\corrl(w)$  as the residue of $w$.
Aiming to estimate $\hl(u,w)\cdot\corrl(w)\cdot\hrlap(v,w)$ as a whole, we propose \revspread that propagates the residue over graph $G$, following the out-going edges of every encountered node.

In this section, we call the $(\ell-1)$-th level as the next level of the $\ell$-th level.
At current $\ell$-th level, by pushing initial residue $\resil(w)$ to the out-neighbors $v$ of $w$ in $G$, nodes $v$ accumulate residue $\resilminusone(v)$ at the $(\ell-1)$-th level.
Then, \revspread goes to the next level to push.
After  $\ell$ iterations, we have many nonzero $\resilzero(v)$.
Then $\resilzero(v)$ estimates $\hl(u,w)\cdot\corrl(w)\cdot\hrlap(v,w)$ with respect to $\resil(w)$, and thus, $\resilzero(v)$ is added to $\simap(u,v)$.
Figure \ref{fig:revpushexample} shows an example that is already explained in Section \ref{sec:overview}.
Further, instead of independently push for each attention node, we \textit{combine} the push of the residues that are aggregated at the same node at the same level.
For example, given node $w$ with $\resi^{(3)}(w)$ at the $3$-rd level of $\Tsrc$  and $w'$ with $\resi^{(2)}(w')$ at the $2$-nd level,
after pushing $\resi^{(3)}(w)$ to the out-neighbors $v$ of $w$ in $G$, we obtain many $\resi^{(2)}(v)$. If $w'$ happens to be an out-neighbor of $w$, the residue that it gets from $w$ and the residue of itself $\resi^{(2)}(w')$ are combined and pushed together.

Algorithm \ref{alg:revspread} shows the pseudo code of \revspreadnospace, which returns the estimated single source SimRank values.
At Line 1, SimRank values $\simap(u,v)$ are initialized to zeros for $v\in V$.
At Line 2, the initial residue of each attention node $w$ is $\resil(w)$, and  
the residues of all other nodes at all levels  are zeros by default.
Starting from level $\ell'=L$ to $1$, for every node $v'$ with residue $\resilprime(v^\prime)$ that satisfies $\sqrtc\cdot\resilprime(v^\prime)\geq\epsp$, we propagate its residue to its out-neighbors $v$ (Lines 3-5).
If $\ell'>1$, residue $\resilprimeminusone(v)$ is increased by  $\frac{\sqrtc\cdot \resilprime(v^\prime)}{\inD(v)}$, where $\inD(v)$ is the indegree of $v$ in $G$ (Lines 6-7); if $\ell'$ is $1$, which means $v$ is at the $0$-th level,  $\simap(u,v)$ is increased by $\frac{\sqrtc\cdot \resilprime(v^\prime)}{\inD(v)}$ at Line 9. 
Finally, $\simap(u,u)$ is set to 1 and all SimRank values $\simap(u,v)$ for all $v\in V$ are returned at Lines 10-11.
The time complexity of Algorithm \ref{alg:revspread} is presented in Lemma \ref{lemma:revspreadcomplex}.

\begin{lemma}\label{lemma:revspreadcomplex}
Algorithm \ref{alg:revspread} runs in $O(\complexrevpush)$  time.
\end{lemma}

\begin{algorithm}[!t]
\begin{small}
\caption{$\mathsf{\revspread}$}\label{alg:revspread}
\BlankLine
\KwIn{Residues $\resil(w)$ of all attention nodes}
\KwOut{$\simap(u,v)$ for $v\in V$}
$\simap(u,v)\gets 0$ for $v\in V$\;
$\resilprime(v)\gets 0$, for $\ell'=1,...,L$ and $v\in V$, except the initial residues $\resil(w)$ of all attention nodes $w$\;
\For{$\ell' = L, ..., 1$}{
    \For{each $v'$ with  $\sqrtc\cdot\resilprime(v')\geq\epsp$}{
        \For{each $v\in \outN(v')$}{
            \If{$\ell'-1>0$}{
            $\resilprimeminusone(v)\gets \resilprimeminusone(v) + \frac{\sqrtc\cdot \resilprime(v')}{\inD(v)}$\;
            }\Else{
            $\simap(u,v)\gets \simap(u,v) + \frac{\sqrtc\cdot \resilprime(v')}{\inD(v)}$
            }
        }
    }
}
$\simap(u,u)\gets 1$\;
\Return $\simap(u,v)$ for $v\in V$\;
\end{small}
\end{algorithm}

\begin{figure*}[!t]
\centering
\begin{small}
\begin{tabular}{cccc}
\multicolumn{4}{c}{\hspace{-2mm} \includegraphics[height=2.8mm]{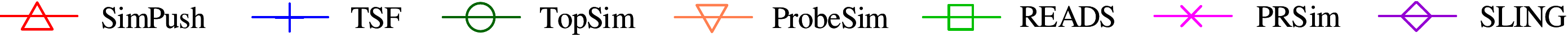}}  \\[-1mm]
\hspace{-4mm} \includegraphics[height=32mm]{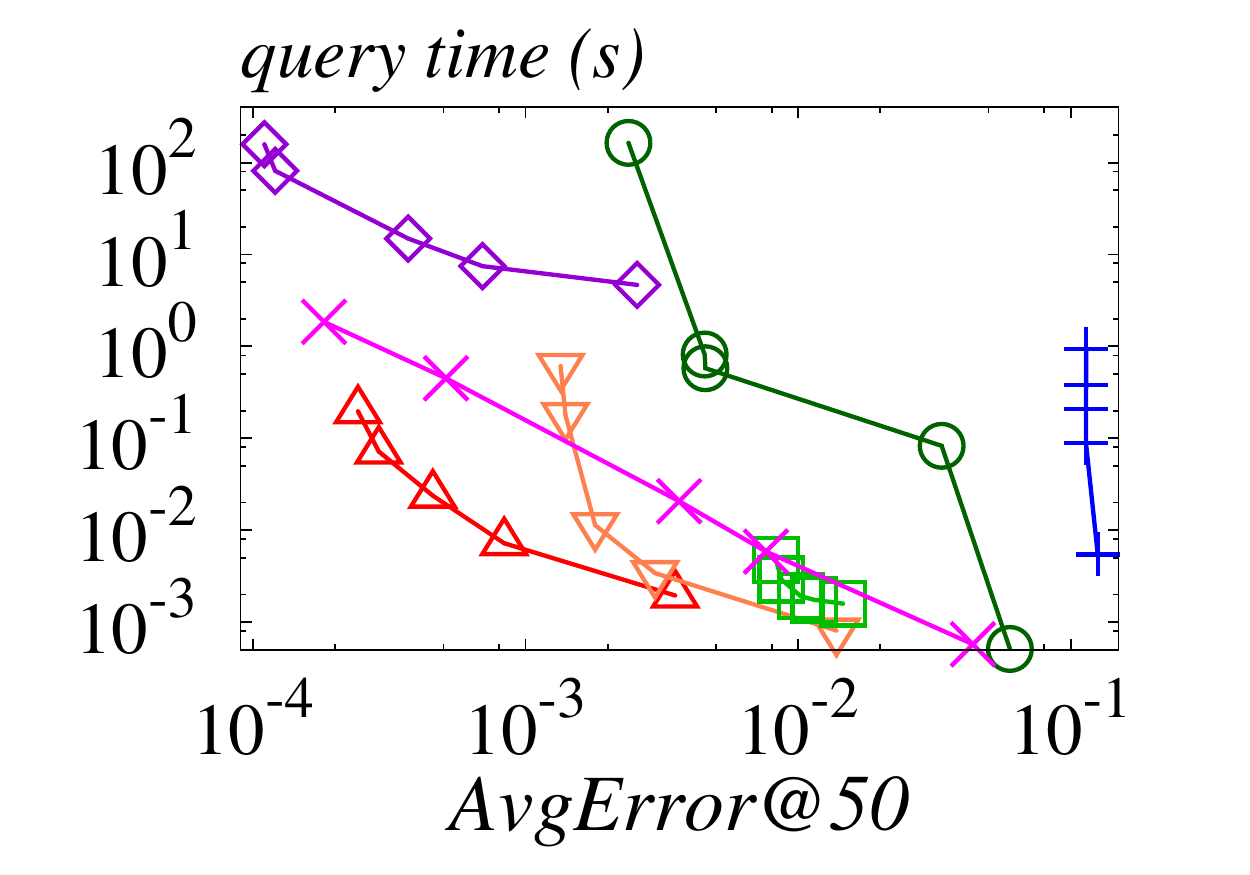}&
\hspace{-4mm} \includegraphics[height=32mm]{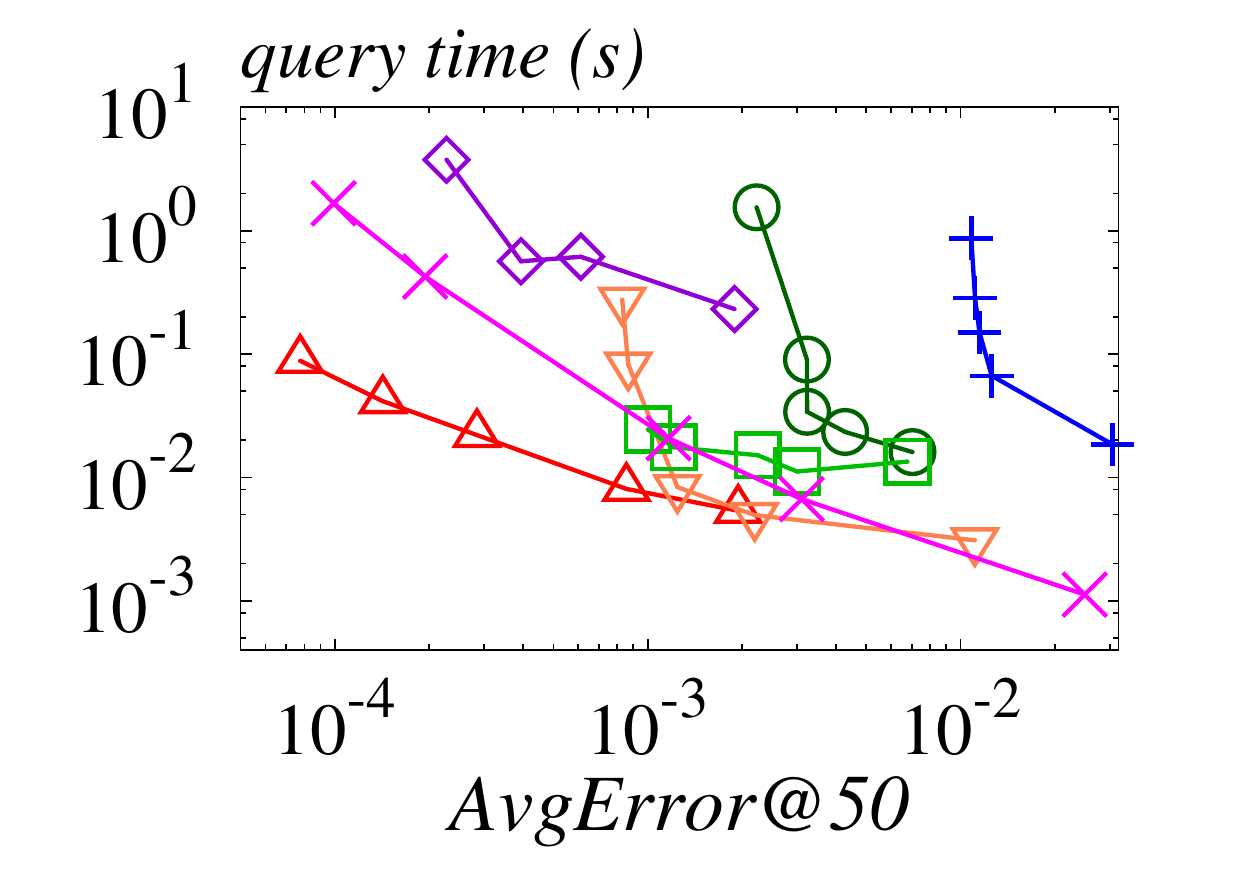} &
\hspace{-4mm} \includegraphics[height=32mm]{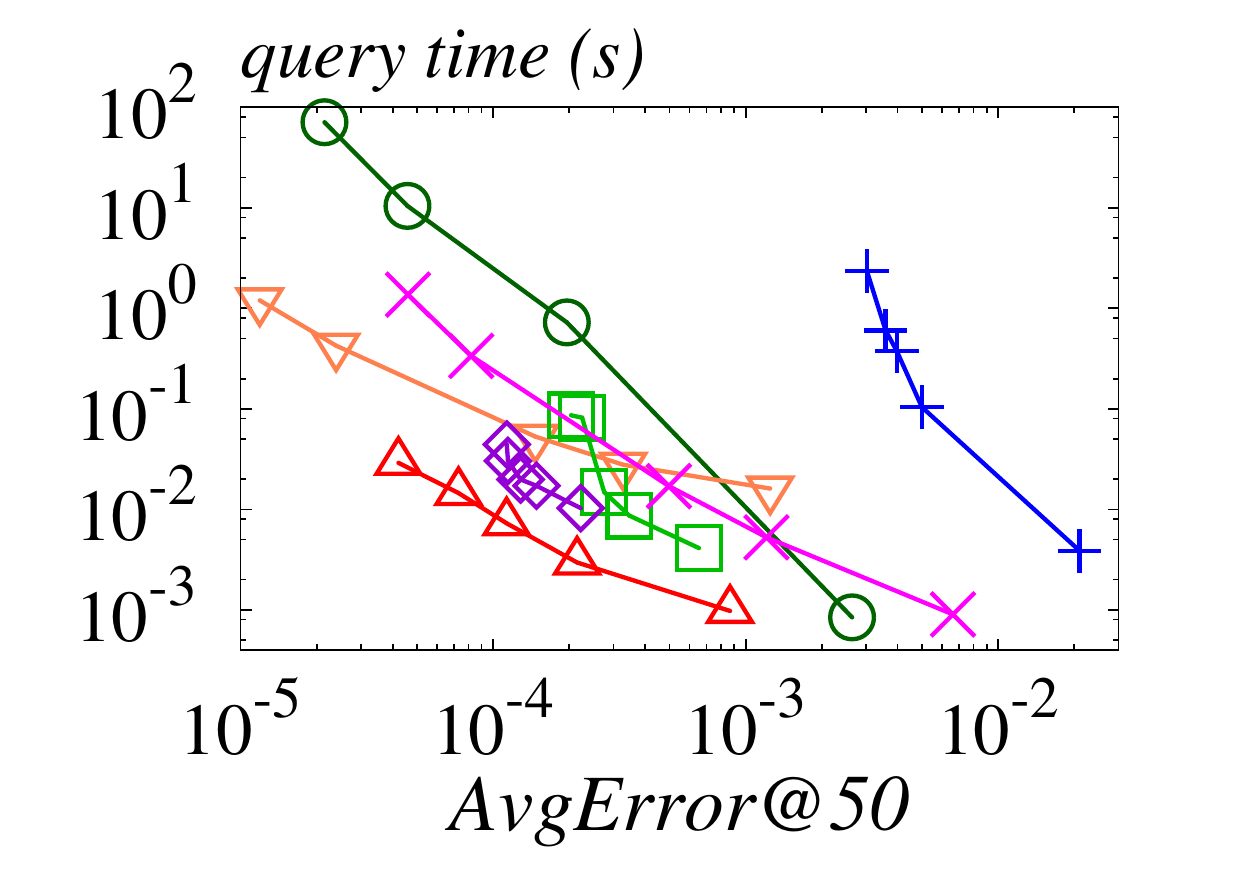}&
\hspace{-4mm} \includegraphics[height=32mm]{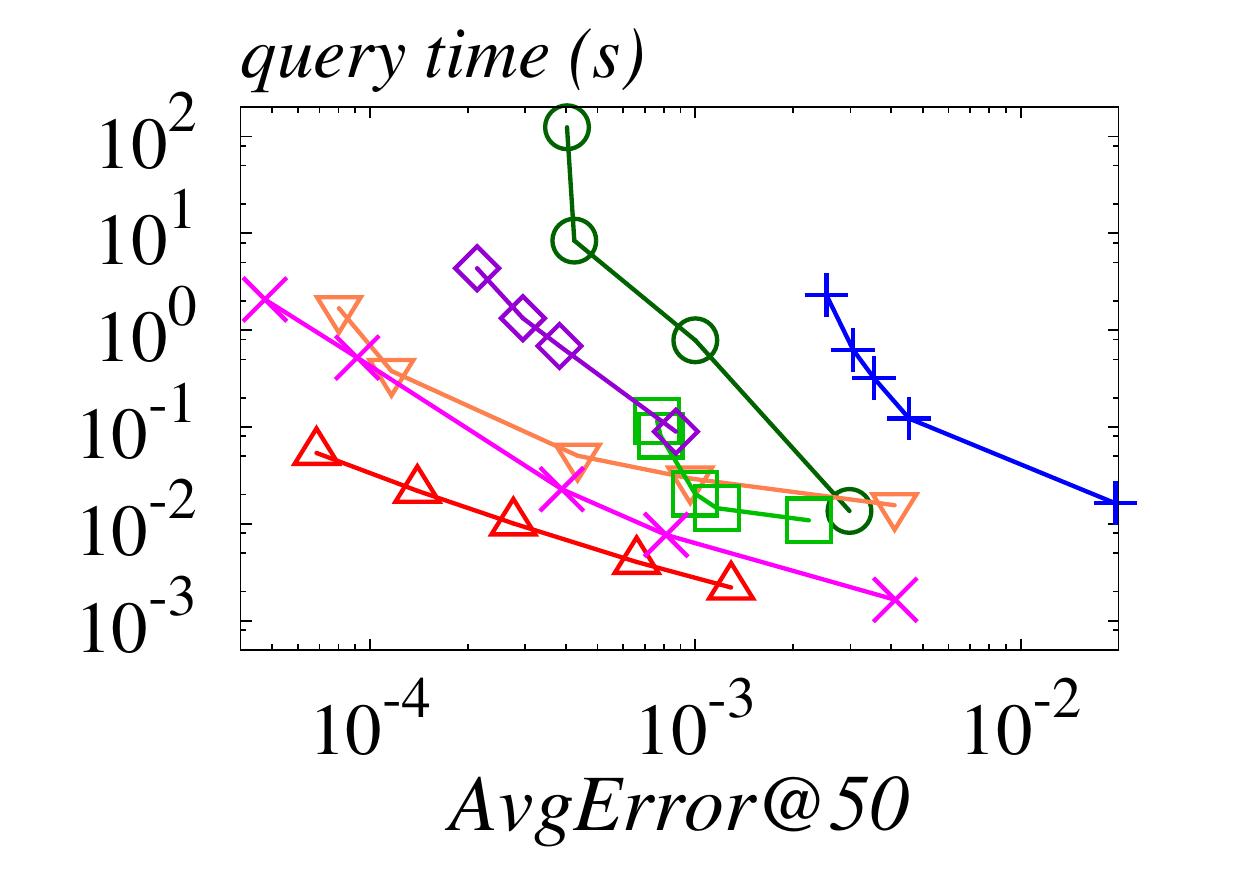} 
\\[-1mm]
\hspace{-4mm} \revise{(a) In-2004} &
\hspace{-4mm} \revise{(b) DBLP}  &
\hspace{-4mm} \revise{(c) Pokec} &
\hspace{-4mm} \revise{(d) LiveJournal}
\\[0mm]
 \hspace{-4mm} \includegraphics[height=32mm]{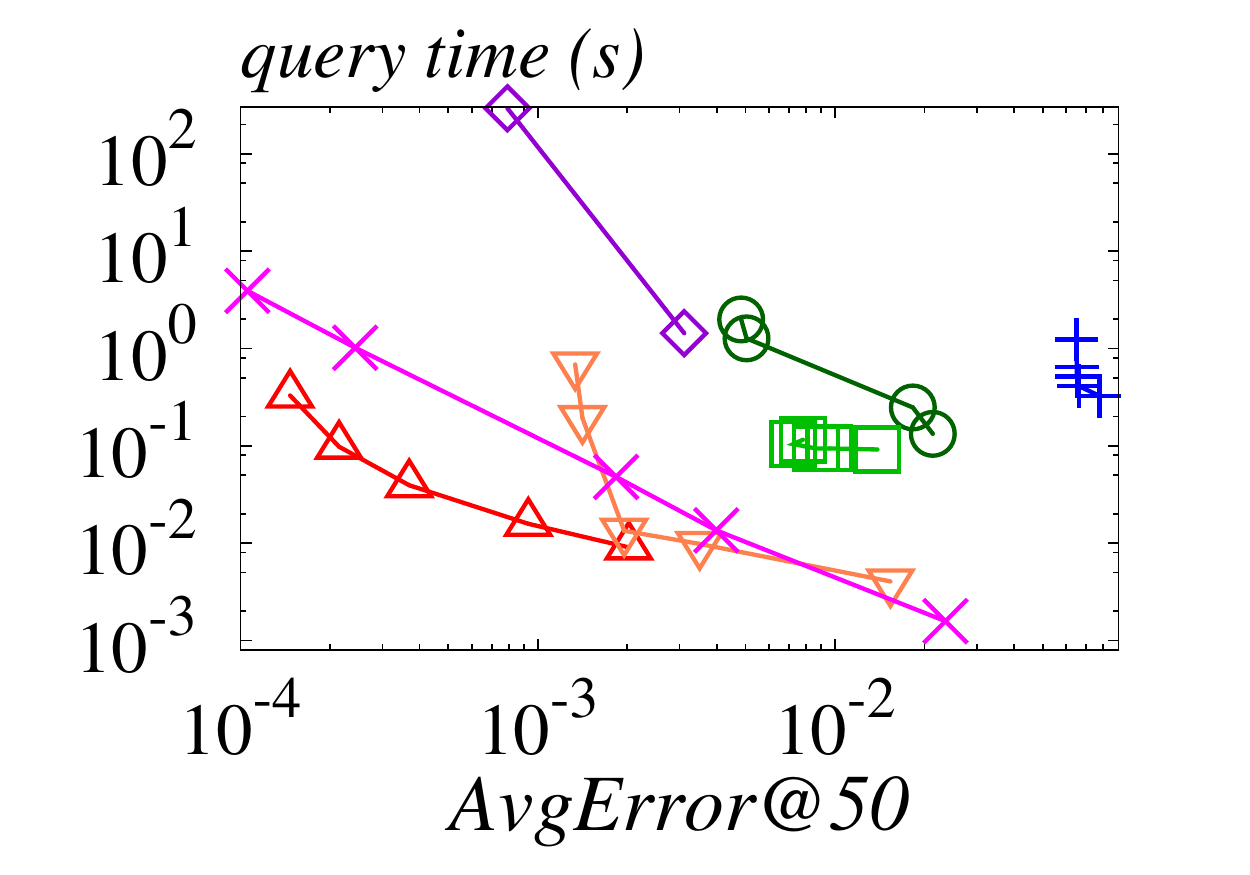} &
\hspace{-4mm} \includegraphics[height=32mm]{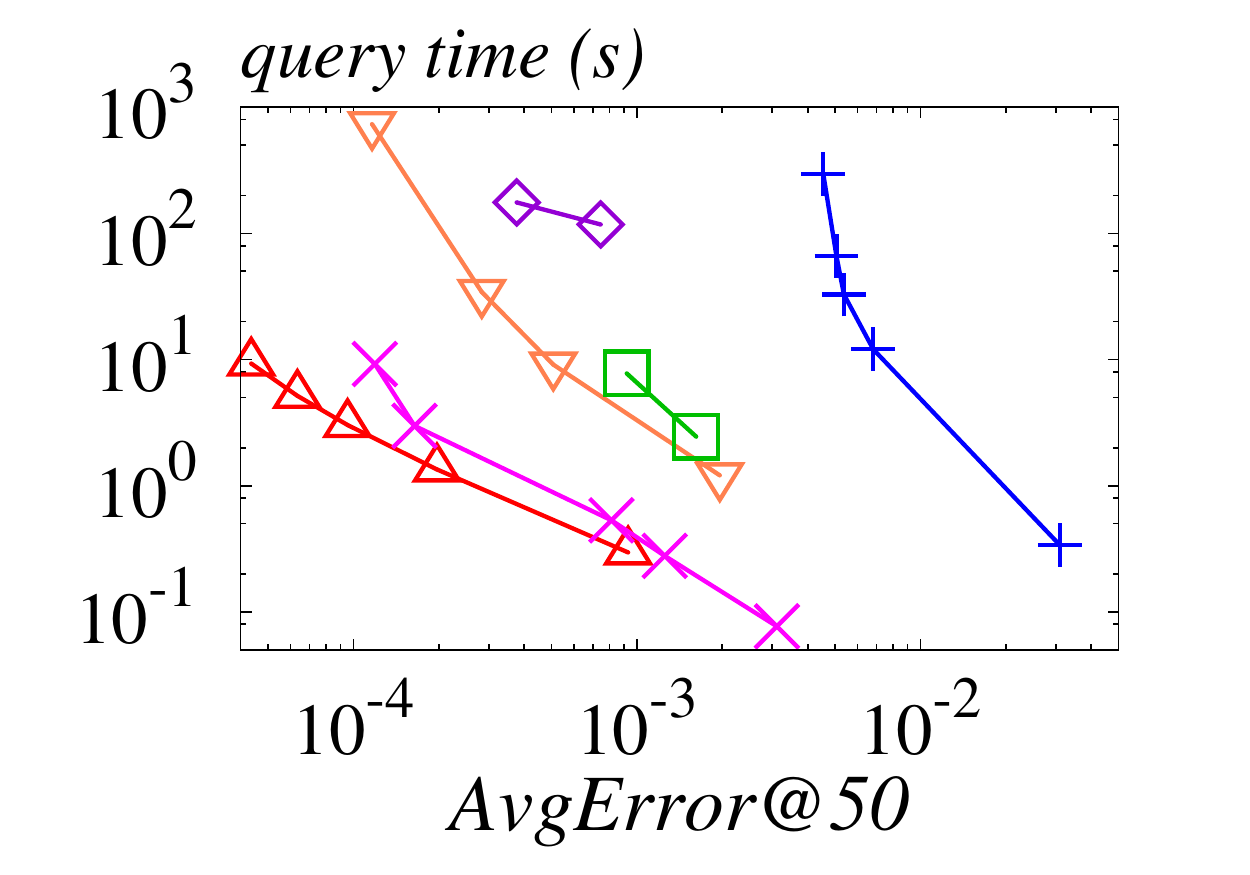}&
\hspace{-4mm} \includegraphics[height=32mm]{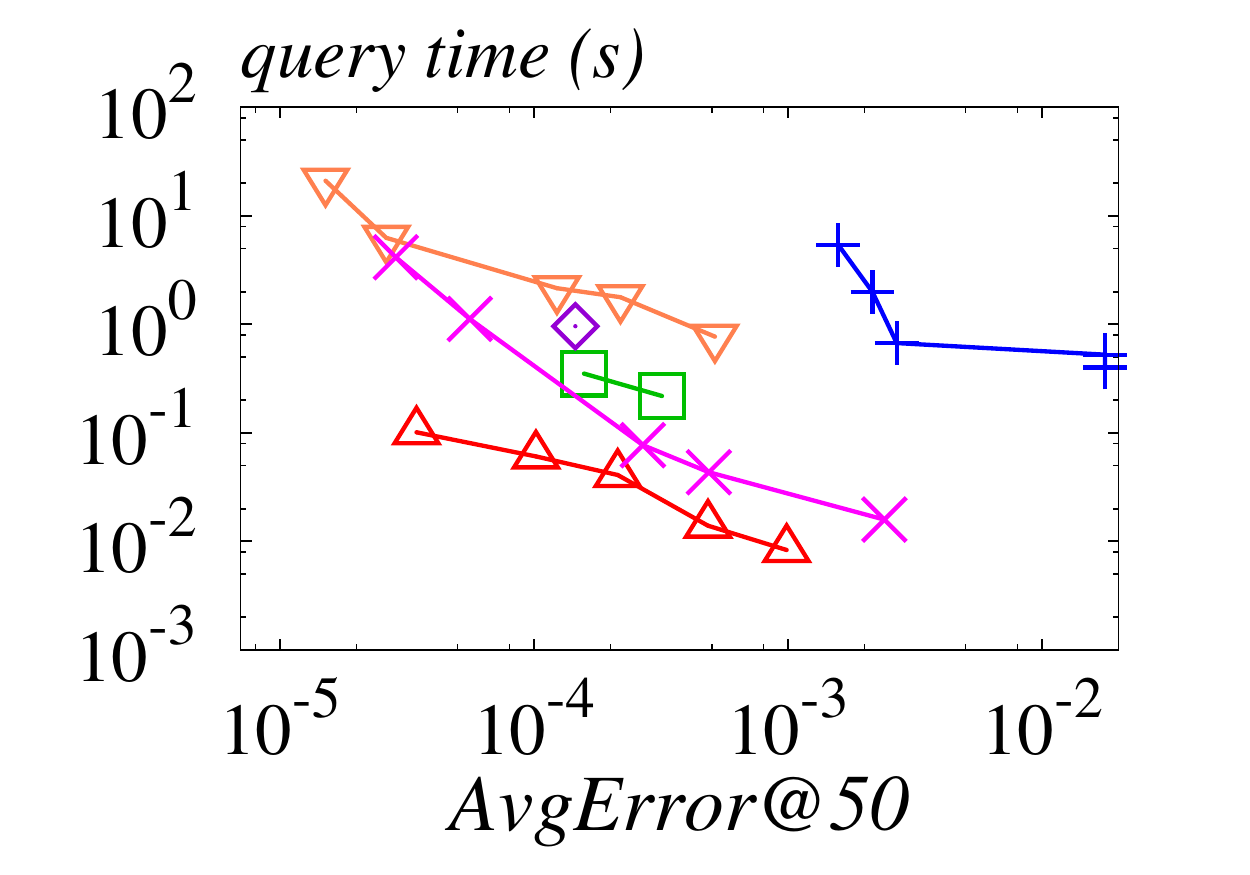} &
\hspace{-4mm} \includegraphics[height=32mm]{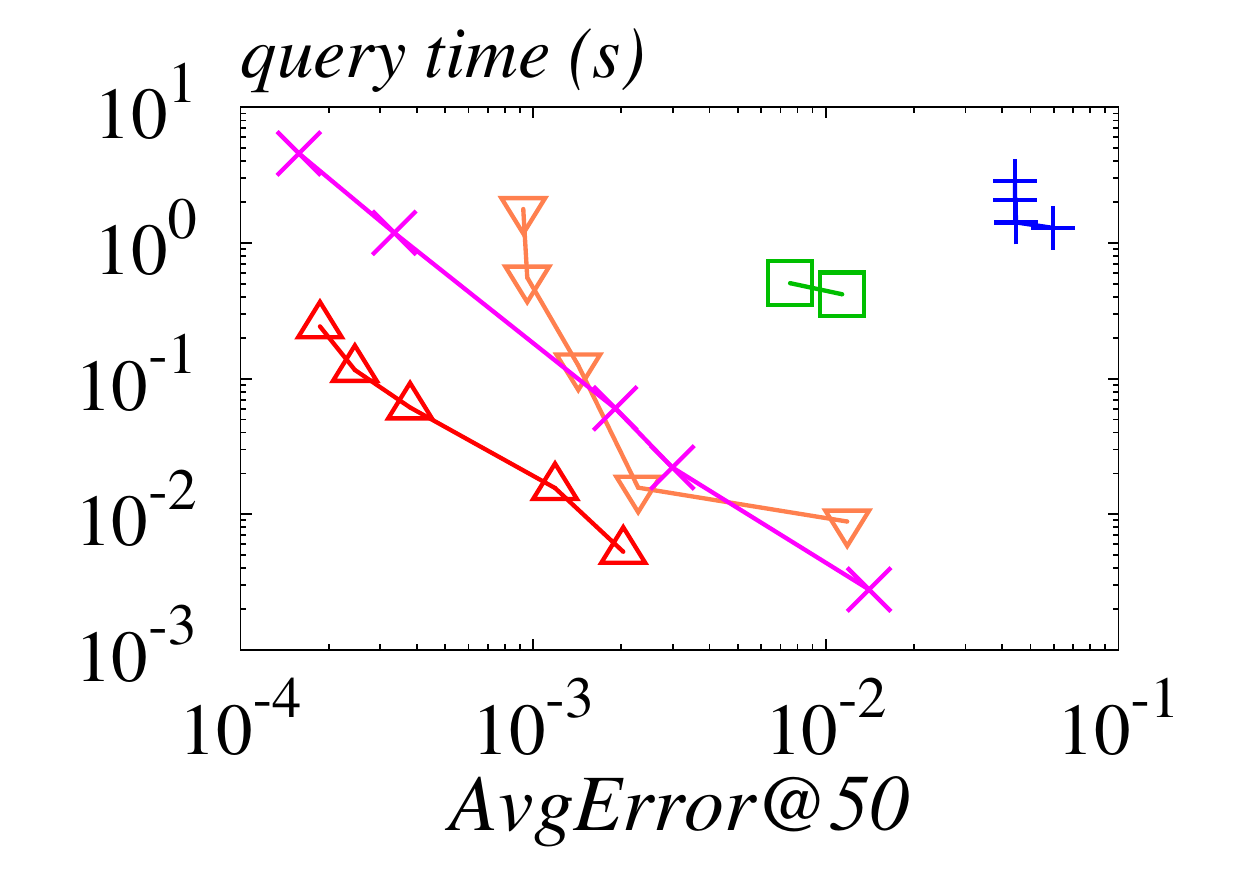}
\\[-1mm]
\hspace{-4mm} \revise{(e)  IT-2004} &
\hspace{-4mm} \revise{(f) Twitter} &
\hspace{-4mm} \revise{(g) Friendster} &
\hspace{-4mm} \revise{(h) UK} \\[-1mm]
\end{tabular}
\vspace{-3mm}
\caption{\revise{Average error {\it vs.} Query time}} \label{exp:err-runtime}
\vspace{-3mm}
\end{small}
\end{figure*}

\vspace{-2mm}
\subsection{Correctness and Complexity Analysis}\label{sec:correctness}
Theorems \ref{theorem:correctness} and \ref{theorem:timecomplex} present \ours's accuracy guarantee and time complexity, respectively. 

\begin{theorem}\label{theorem:correctness}
Given graph $G$, query node $u$, error parameter $\eps$, and failure probability $\delta$, Algorithm \ref{algo:main} returns an estimated SimRank value $\simap(u,v)$ that satisfies $s(u,v)-\simap(u,v)\leq \eps$ for each node $v$ in G, with at least $1-\delta$ probability, where $s(u,v)$ is the exact SimRank value between $u$ and $v$.
\end{theorem}


\begin{theorem}\label{theorem:timecomplex}
In expectation, Algorithm~\ref{algo:main} runs in $O(m\log\frac{1}{\eps}/\eps + \log\frac{1}{\eps\delta}/\eps^2 + 1/\eps^3)$ time.
\end{theorem}

\section{Experiments}\label{sec:experiments}

We  evaluate \ours against the state of the art.
All experiments are conducted on a Linux server with an Intel Xeon 2.60GHz CPU and 376GB RAM. All methods are  in C++ and compiled by g++ 7.4 with -O3 optimization.

\subsection{Experimental Settings}\label{sec:exp-set}

\noindent
{\bf Methods.}
\ours is compared with six methods:
\prsim~\cite{prsimWeiHX0LDW19}, \reads~\cite{readsJiangFWW17}, \topsim~\cite{topsimLeeLY12}, \sling~\cite{slingTianX16}, \probesim~\cite{probsimLiuZHWXZL17}, and \tsf\cite{tsfShaoC0LX15}.
\probesim and \topsim are index-free;
\prsim, \reads, \tsf,  \sling are index-based.

\header
{\bf Datasets and query sets.} 
\revise{We use 9 real-world graphs to evaluate \ours and the  competitors.}
\revise{The largest graph, ClueWeb, contains 1.68 billion nodes and 7.94 billion edges.}
The statistics of the graphs are shown in Table \ref{tbl:exp-data}. 
There are \revise{5} large graphs with billions of edges: 
\revise{ClueWeb}, 
UK, Friendster, Twitter, and IT-2004, and 4 smaller graphs with millions of edges: In-2004, DBLP, Pokec, and LiveJournal.
These graphs are of various types, including social networks, web graphs, and collaboration network. 
All datasets are  available at \cite{snapnets,unimidataset, clueweb09}.
\revise{For each dataset, we generate 100 queries by selecting nodes uniformly at random.}

\begin{table}[!t]
\centering
\begin{small}
\caption{Datasets used in the experiments.} \label{tbl:exp-data}
\vspace{-3mm}
\renewcommand{\arraystretch}{1.2}
\begin{tabular}{|l|r|r|r|c|}
    \hline
    {\bf Name} & \multicolumn{1}{c|}{$\boldsymbol{n}$} & \multicolumn{1}{c|}{$\boldsymbol{m}$} & \multicolumn{1}{c|}{\bf Type}  \\
    \hline
    {\em In-2004}   & 1,382,908 & 16,539,643 & directed \\
    \hline
    {\em DBLP}   & 5,425,963 & 17,298,032 & undirected \\
    \hline
    {\em Pokec}   & 1,632,803 & 30,622,564 & directed \\
    \hline
    {\em LiveJournal} &4,847,571 & 68,475,391 & directed  \\
    \hline
    {\em IT-2004}   &41,291,594& 1,135,718,909 & directed \\
    \hline
    {\em Twitter}   &41,652,230 & 1,468,364,884 & directed \\
    \hline
    {\em Friendster}   &65,608,366 & 3,612,134,270 & undirected \\
    \hline
    {\em UK}   &133,633,040 &5,475,109,924 & directed \\
    \hline
    \revise{{\em ClueWeb}}   & \revise{1,684,868,322} & \revise{7,939,635,651} & \revise{directed} \\
    \hline
\end{tabular}
\vspace{-2mm}
\end{small}
\vspace{-4mm}
\end{table}

\header
\textbf{Parameters.}
Following \cite{prsimWeiHX0LDW19,probsimLiuZHWXZL17,slingTianX16}, we set the decay factor $c$ to $0.6$, and fix the failure probability  $\delta=0.0001$.
For \ours, we vary $\eps$ in $\{0.05, 0.02, 	0.01, 0.005, 0.002\}$.  
\revise{We set the parameters of all competitors following the settings in \cite{prsimWeiHX0LDW19}}.
In particular, \prsim has two parameters: $\epsilon_a$, the absolute error threshold, and $j_0$, the number of hub nodes. We vary $\epsilon_a$ in $\{0.5, 0.1, 0.05, 0.01, 0.005\}$, and set $j_0$ to $\sqrt{n}$ by default~\cite{prsimWeiHX0LDW19}. 
We evaluate the static version of \reads, which is the fastest among the three algorithms proposed in \cite{readsJiangFWW17}.
\reads has two parameters: $r$, the number of 
$\sqrt{c}$-walks generated for each node in preprocessing stage, and $t$, the maximum depth of the $\sqrt{c}$-walks.
We vary $(r,t)$ in $\{(10, 2), (50, 5), (100, 10), (500, 10), (1000, 20)\}$.
\topsim has four parameters: $T$, the depth of random walks; $1/h$, the minimal degree threshold to identify a high degree node; $\eta$, the similarity threshold for trimming a random walk; $H$, the number of random walks to be expanded at each level. We fix $H$ and $\eta$ to their default values, {\it i.e.}, $100$ and $0.001$, respectively. We vary $(T,1/h)$ in $\{(1, 10), (3, 100), (3, 1000),(3, 10000), (4, 10000)\}$.
\sling has a parameter $\epsilon_a$, which denotes the upper bound on the absolute error. We vary $\epsilon_a$ in $\{0.5, 0.1, 0.05, 0.01, 0.005\}$.
\probesim also has an absolute error threshold $\epsilon_a$, which we vary in $\{0.5, 0.1, 0.05, 0.01, 0.005\}$.
\tsf has two parameters $R_g$ and $R_q$, which are the number of one-way graphs stored in the index and the times each one-way graph reused during query processing, respectively. We vary  $(R_g,R_q)$ in $\{(10,2),(100,20),(200,30),(300,40),(600,80)\}$. 
\revise{Note that every method is evaluated using its respective five parameter settings listed above; for each method, from its first to last parameter settings, it generates increasingly accurate results, with higher running time and memory usage.}

\begin{figure*}[!t]
\centering
\begin{small}
\begin{tabular}{cccc}
\multicolumn{4}{c}{\hspace{-2mm} \includegraphics[height=2.8mm]{./figure/algo-legend.pdf}}  \\[-1mm]
\hspace{-4mm} \includegraphics[height=32mm]{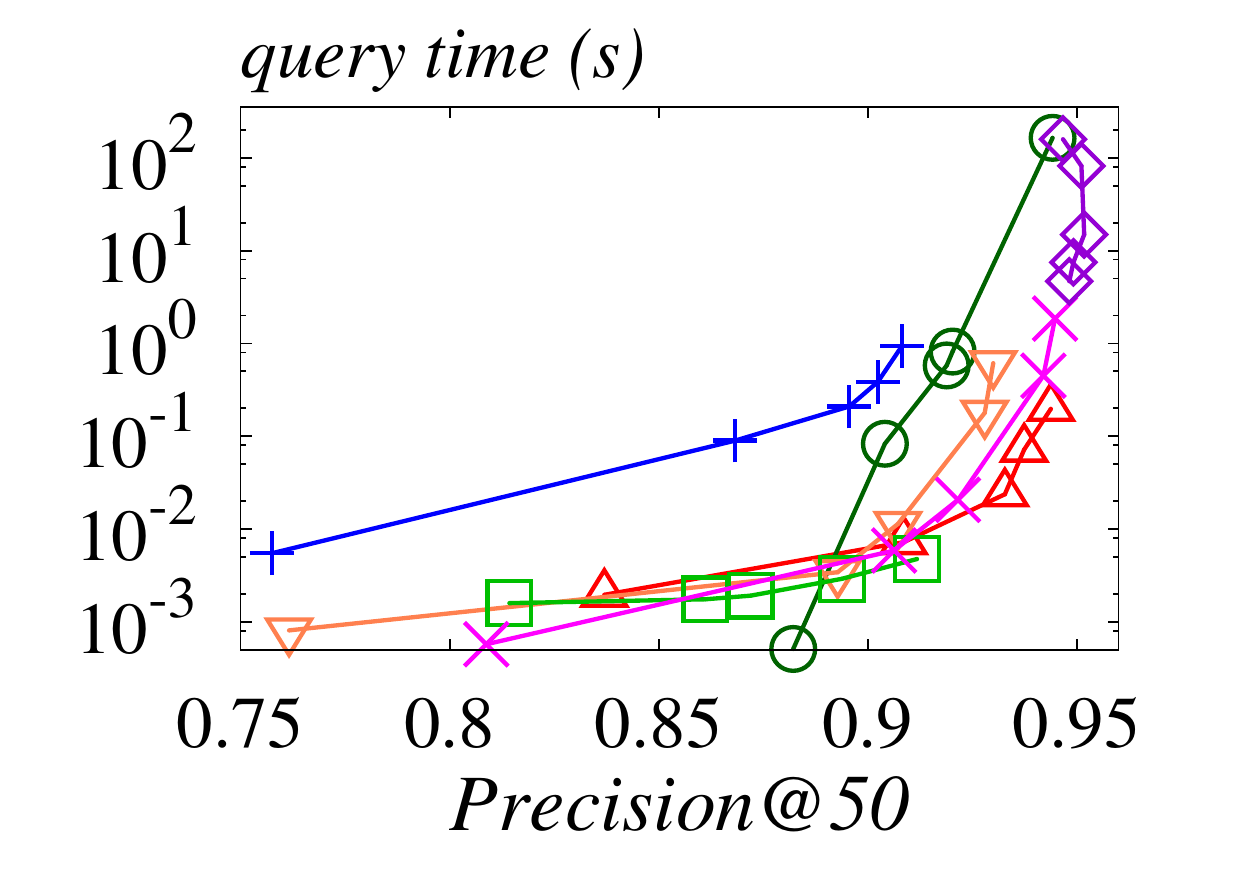} &
\hspace{-4mm} \includegraphics[height=32mm]{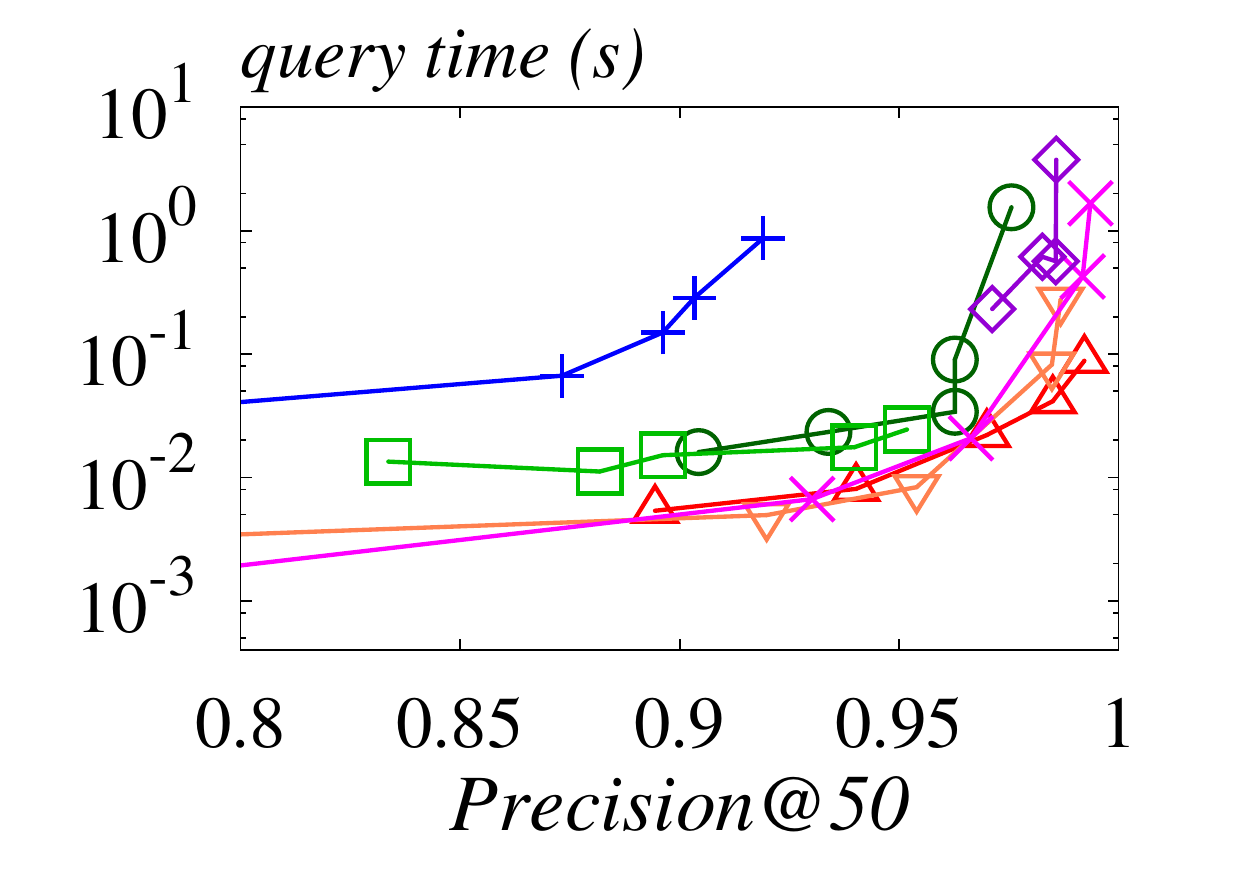} &
\hspace{-4mm} \includegraphics[height=32mm]{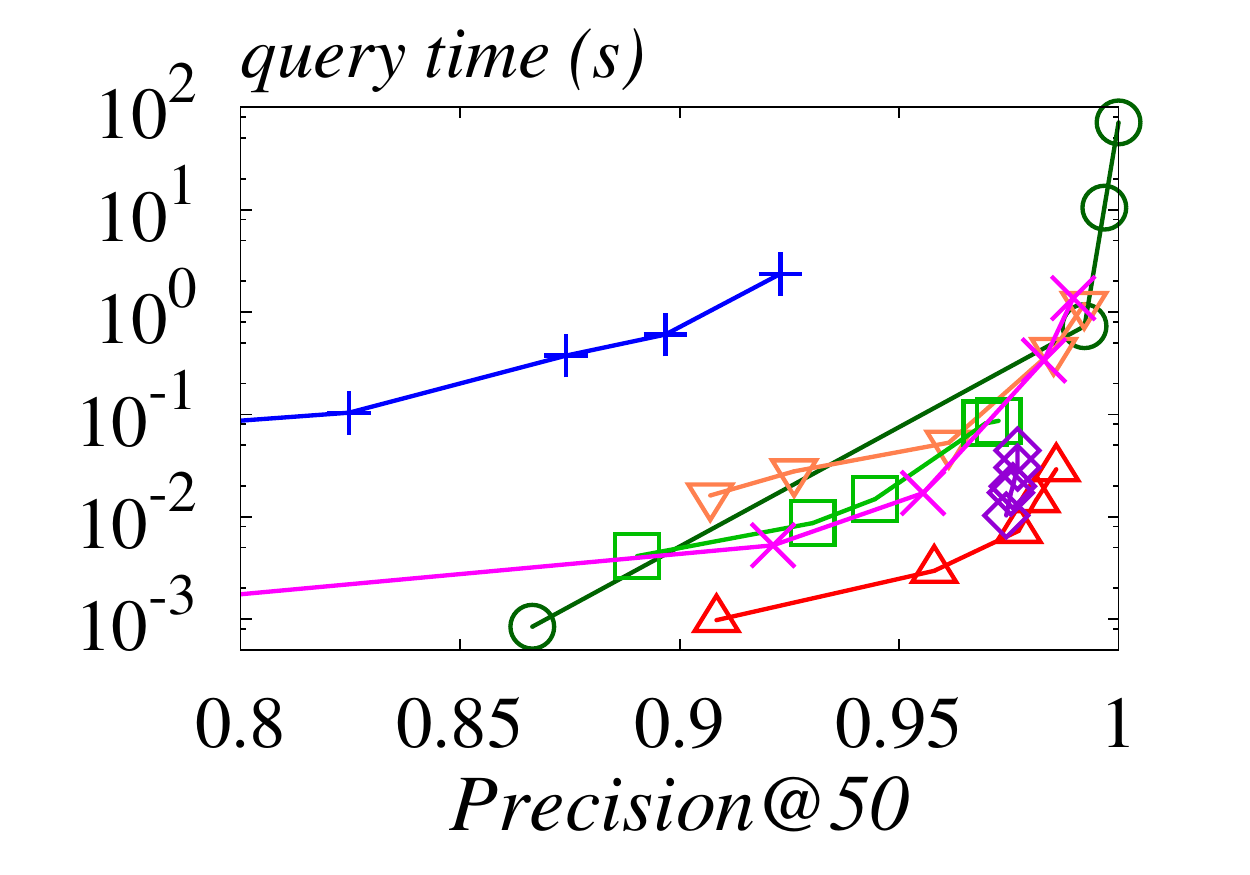}&
\hspace{-4mm} \includegraphics[height=32mm]{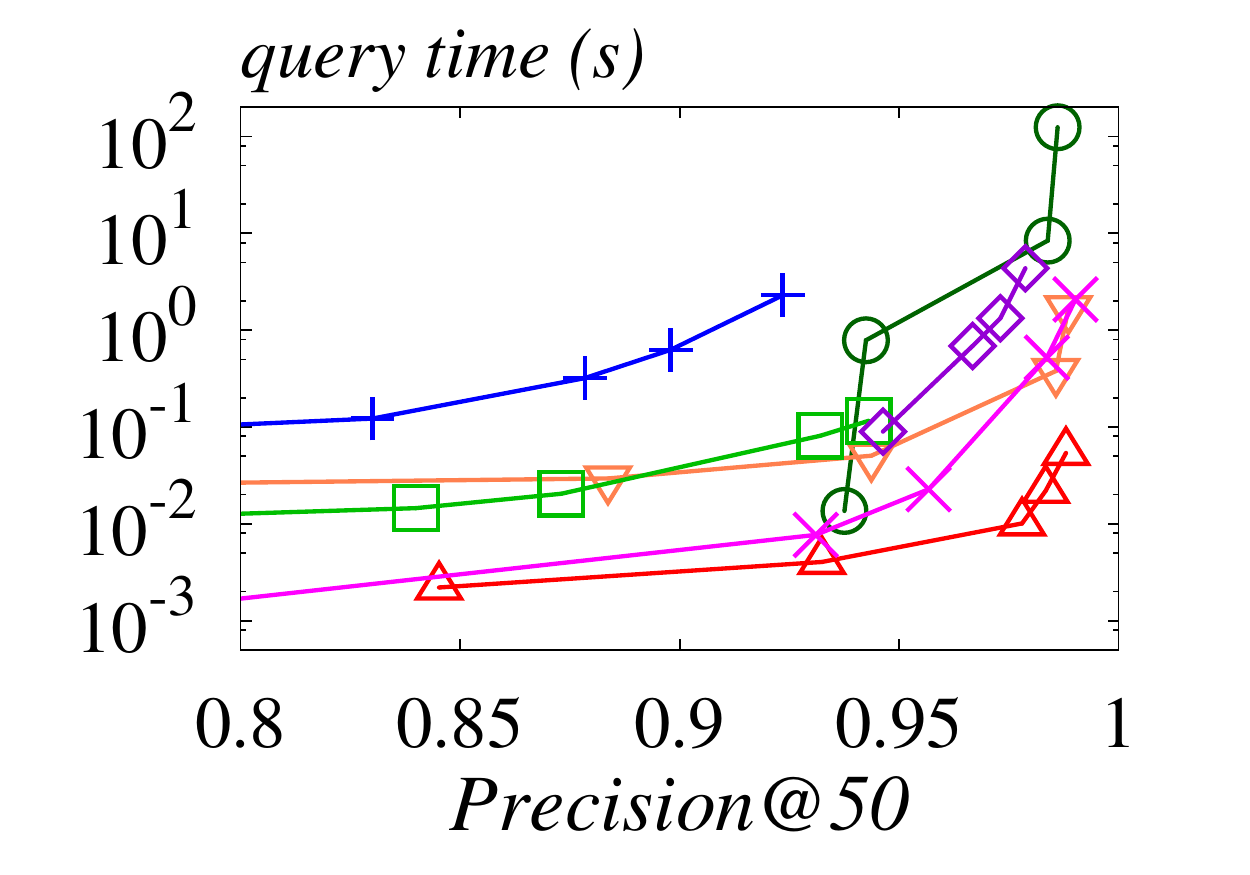} 
\\[-1mm]
\hspace{-4mm} \revise{(a) In-2004} &
\hspace{-4mm} \revise{(b) DBLP}  &
\hspace{-4mm} \revise{(c) Pokec} &
\hspace{-4mm} \revise{(d) LiveJournal} 
\\[0mm]
 \hspace{-4mm} \includegraphics[height=32mm]{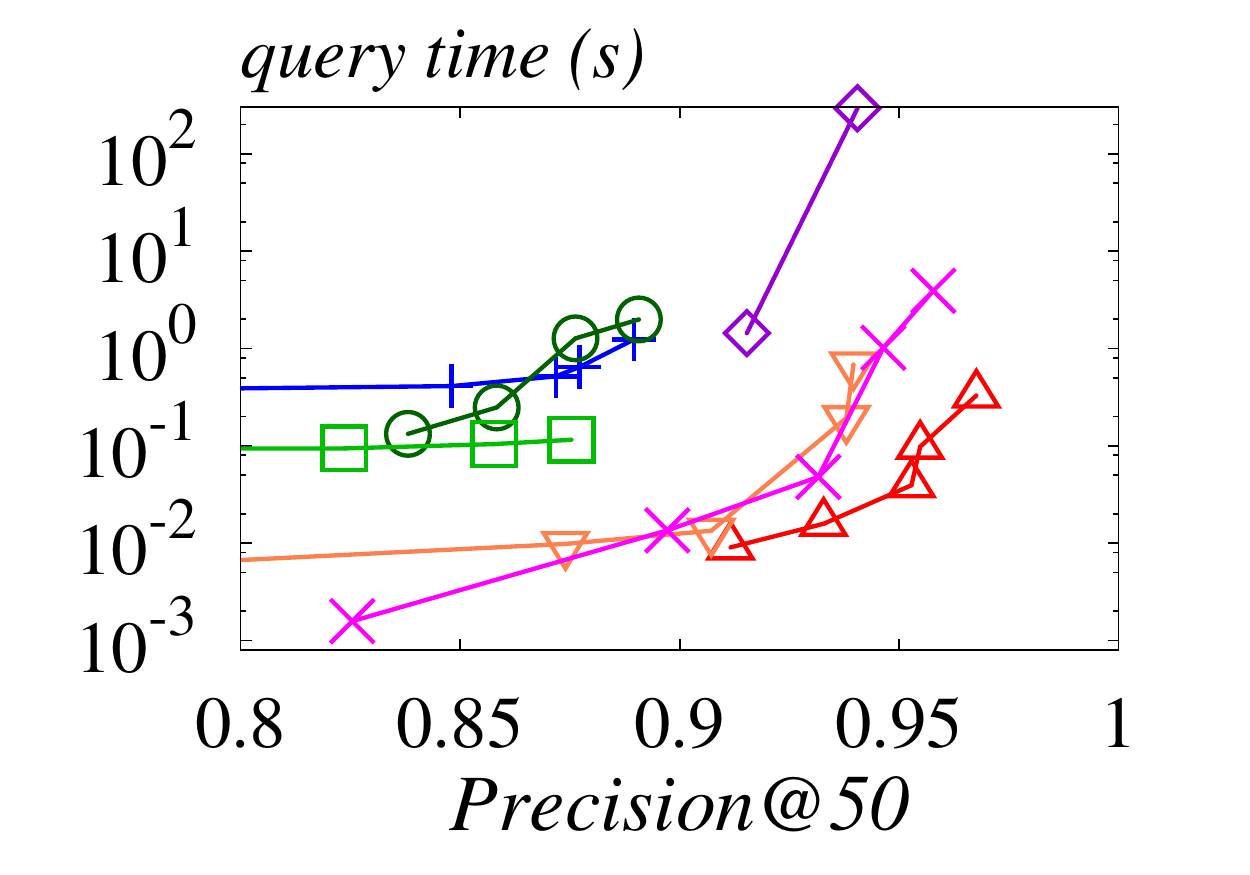} &
\hspace{-4mm} \includegraphics[height=32mm]{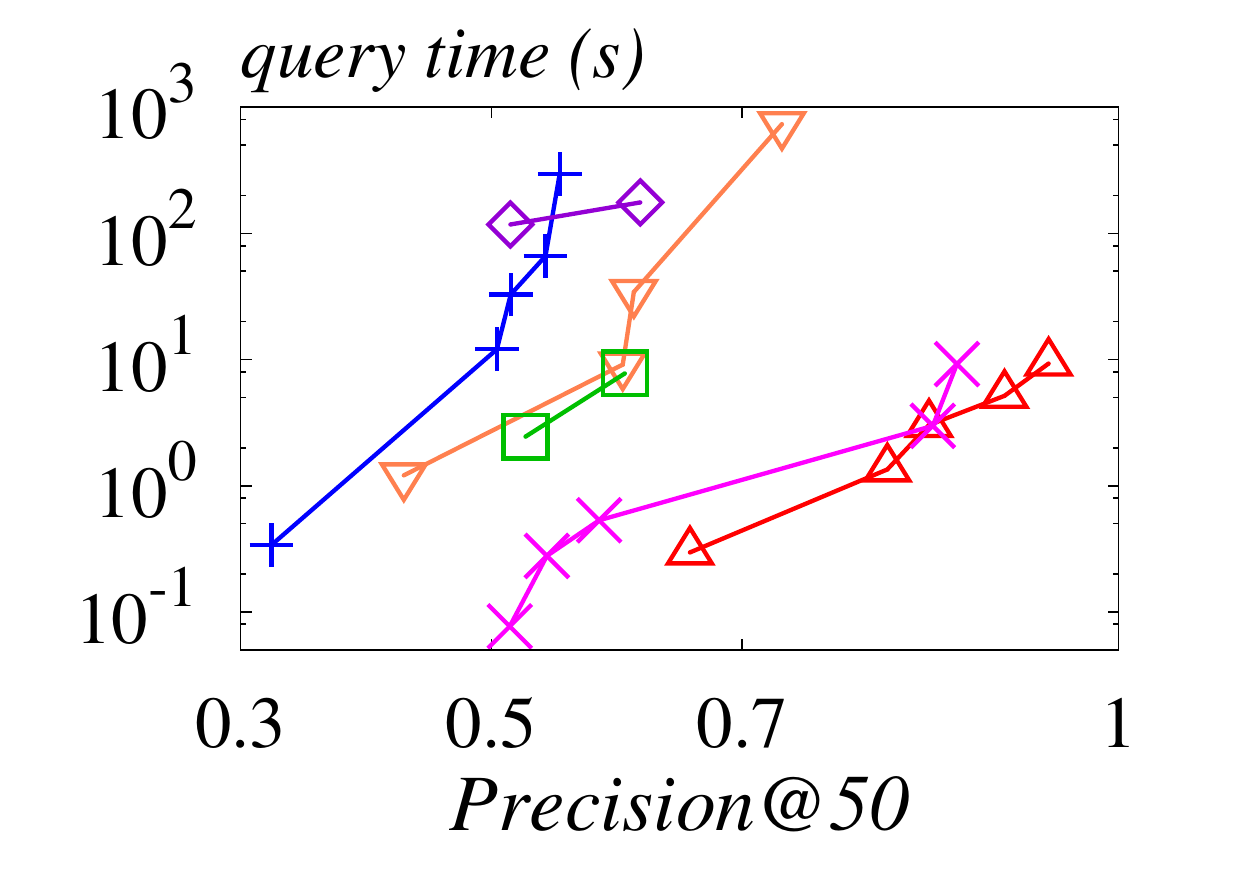}&
\hspace{-4mm} \includegraphics[height=32mm]{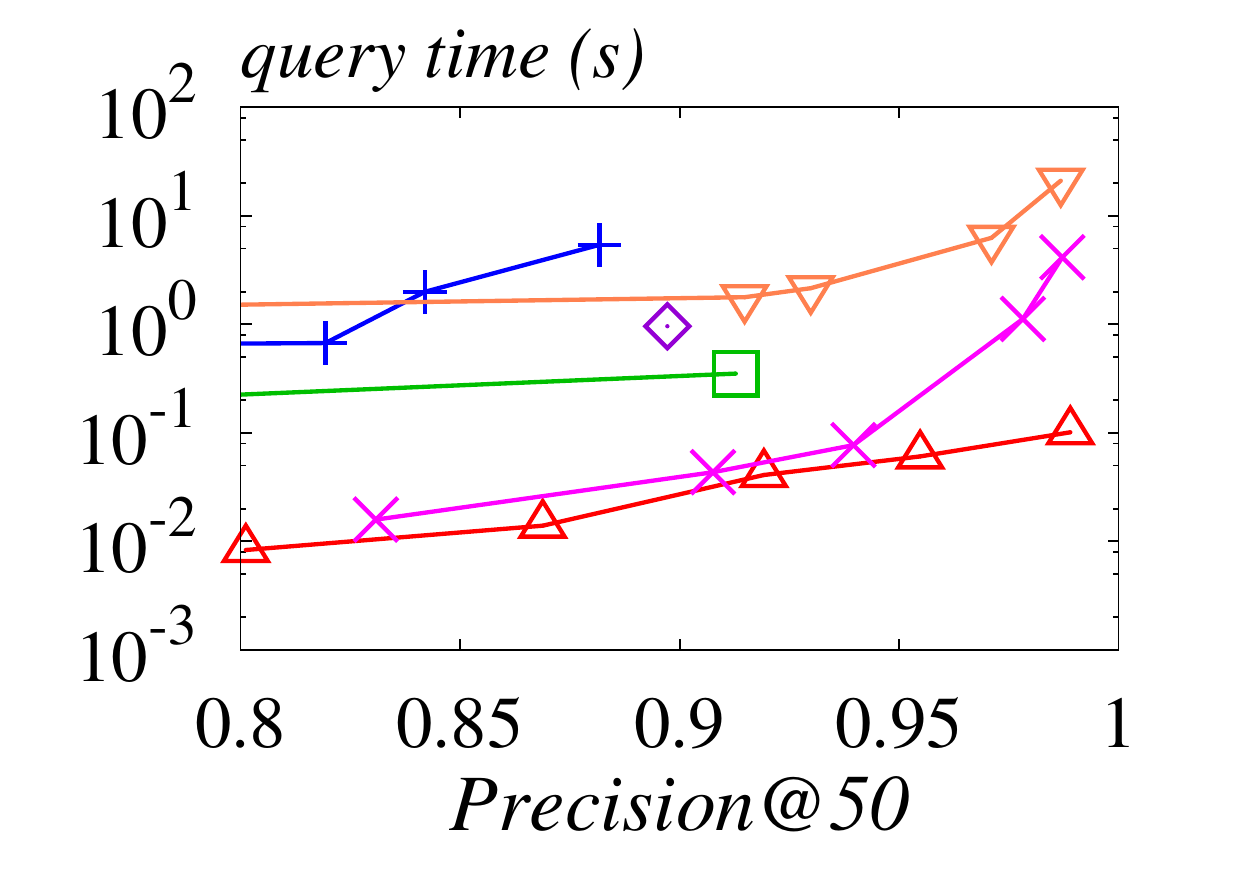} &
\hspace{-4mm} \includegraphics[height=32mm]{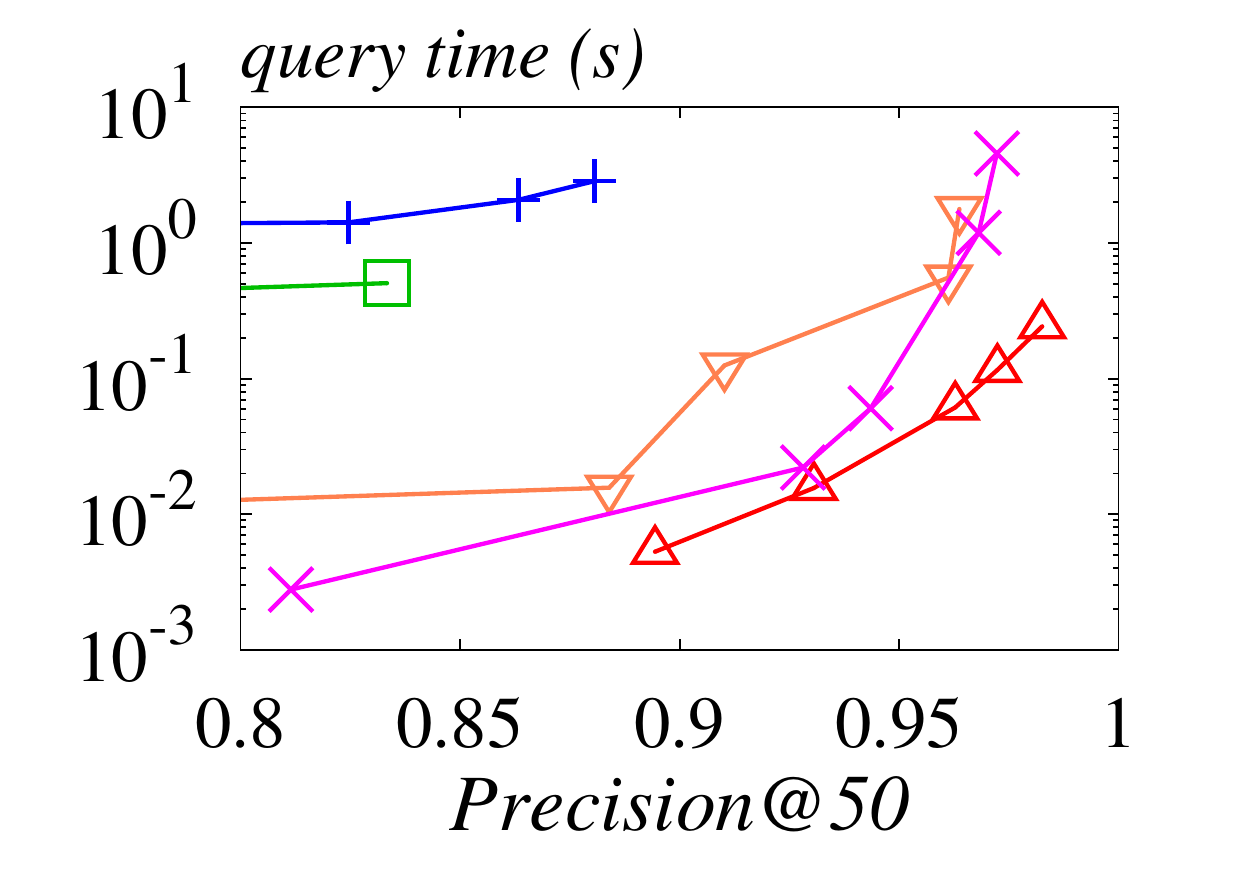}
\\[-1mm]
\hspace{-4mm} \revise{(e) IT-2004} &
\hspace{-4mm} \revise{(f) Twitter} &
\hspace{-4mm} \revise{(g) Friendster} &
\hspace{-4mm} \revise{(h) UK} \\[-1mm]
\end{tabular}
\vspace{-2mm}
\caption{\revise{Precision {\it vs.} query time.}} \label{exp:prec-runtime}
\vspace{-3mm}
\end{small}
\end{figure*}

\header
{\bf Ground truth.} 
We get  ground truth for the queries of all datasets by adopting the methods in  \cite{prsimWeiHX0LDW19,probsimLiuZHWXZL17}. 
For small graphs, we directly apply Monte Carlo  \cite{fogaras2005scaling} to estimate SimRank for each query  $u$ and each $v$ in $G$ with an absolute error less than $0.000001$ and confidence over $99.999\%$, which is then used as the ground truth for  $s(u,v)$.
For large graphs, it is expensive to directly employ  Monte Carlo  to estimate $s(u,v)$  with high precision. We adopt the {\em pooling} method \cite{probsimLiuZHWXZL17,prsimWeiHX0LDW19} to generate ground truth for large graphs.
Given query node $u$, we run each single-source algorithm, merge the top-$k$ nodes of each algorithm, remove the duplicates, and put them into a pool. For each node $v$ in the pool, we obtain the ground truth of $s(u,v)$ by  Monte Carlo. The ground truth top-$k$ node set $V_k$ is then the set of $k$ nodes with highest estimated SimRank values from the pool.

\header
{\bf Metrics.} We adopt two metrics for accuracy evaluation, {\it i.e.}, {\em AvgError@k} and {\em Precision@k}, which are also used in \cite{prsimWeiHX0LDW19}. We also evaluate the peak memory usage.
{\em AvgError@k} measures the average absolute error for approximating $s(u,v_i)$ for each node $v_i$ in the ground truth top-$k$ nodes $V_k$. Specifically, for each node $v_i$ in $V_k$, let $\hat{s}(u,v_i)$ denote the estimation of exact  $s(u,v_i)$, {\em AvgError@k} is  defined as
$$
\textstyle
AvgError@k = \frac{1}{k}\sum_{1\le i \le k}{|\hat{s}(u,v_i)-s(u,v_i)|}.
$$

{\em Precision@k} evaluates the ability to return the top-$k$ nodes for a query in terms of ground truth top-$k$ node set $V_k$. Suppose that $V'_k=\{v'_1,\cdots,v'_k\}$ is the top-$k$ nodes returned by the algorithm to be evaluated. {\em Precision@k} is defined as 
$$
Precision@k=|V_k \cap V'_k|/k.
$$
{\em Peak memory usage.} We  enquiry Linux system resource files for \textit{rusage.ru\_maxrss}, to get the peak memory usage of all methods over all datasets under all parameter settings.

\subsection{Experimental Results}\label{sec:exp-result}
We evaluate the tradeoff between average error and query time, the tradeoff between average precision and query time, and the tradeoff between average error and peak memory usage for all methods over all graphs.
We exclude a parameter of a method if it runs out of memory, or cannot finish preprocessing within 24 hours, or cannot finish a query in 1000 seconds.
Given the query set of each graph, for each parameter setting of each method, we report the averages of query time, \textit{AvgError@50}, \textit{Precision@50}, and peak memory usage.
Note that the preprocessing time of the index-based methods are not reported since our method \ours is index-free.

\header
{\bf Average error and query time.}
Figure \ref{exp:err-runtime} reports the tradeoff between {\em AvgError@50} and query time of all methods over the first eight graphs in Table \ref{tbl:exp-data} \revise{(results on ClueWeb are reported separately later on)}.
$x$-axis is  {\em AvgError@50} and $y$-axis is  query time in second(s); both are in log-scale.
\revise{For each method, the plot contains a curve with 5 points, which corresponds to results for its 5 settings (from right to left) described earlier.}
\ours is superior over all methods by achieving lower error with less query time, and consistently outperforms existing solutions, especially on large graphs,  no matter whether the competitor is index-free (e.g., \probesim) or index-based (e.g., \prsim).
To reach the same level of empirical error, \ours is much faster than the competitors, often by over an order of magnitude.
\revise{On  graph UK, in Figure \ref{exp:err-runtime}(h), to achieve \revise{$3.5\times 10^{-4}$} {\em AvgError@50}, \ours uses \revise{$0.062$} seconds, while the index-based state-of-the-art \prsim needs \revise{1.18} seconds, and the index-free \probesim uses \revise{$1.9$} seconds and only achieves $9\times 10^{-4}$ error.}
In Figure \ref{exp:err-runtime}(f) for Twitter, which is known as a hard graph for SimRank computation due to its locally dense structure as analyzed in the paper of \prsim \cite{prsimWeiHX0LDW19}, \ours also outperforms \prsim by a significant gap.
\revise{To achieve \revise{$1.4\times 10^{-4}$} error, \prsim requires $2.7$ hours of precomputation and \revise{$9.1$} seconds for query processing, while our online method \ours  only needs \revise{$1.5$} seconds in total to achieve the same level of error. For \probesim, it needs \revise{$725$} seconds to achieve such error on Twitter.}
As aforementioned, the max level $L$ of  $\Tsrc$ is usually small for real-world graphs. For instance, when $\epsilon=0.02$, on Twitter, $L$ is just $2.76$ on average, and on DBLP, $L$ is $9.0$. This indicates that the attention nodes that can largely contribute to the SimRank values are truly in the vicinity of query nodes.
The number of attention nodes is usually in dozens or hundreds.
Therefore, \ours that first finds the attention nodes that can largely contribute to SimRank values and then focuses on such nodes for estimation, is rather efficient.
On the graphs in Figures \ref{exp:err-runtime}(a)-(d), \ours also exceeds all competitors by a large gap. \sling, \reads, \tsf, and \topsim, are all inferior to \ours over all these graphs.


\begin{figure*}[!t]
\centering
\begin{small}
\begin{tabular}{cccc}
\multicolumn{4}{c}{\hspace{-2mm} \includegraphics[height=2.8mm]{./figure/algo-legend.pdf}}  \\[0mm]
\hspace{-5mm} \includegraphics[height=32mm]{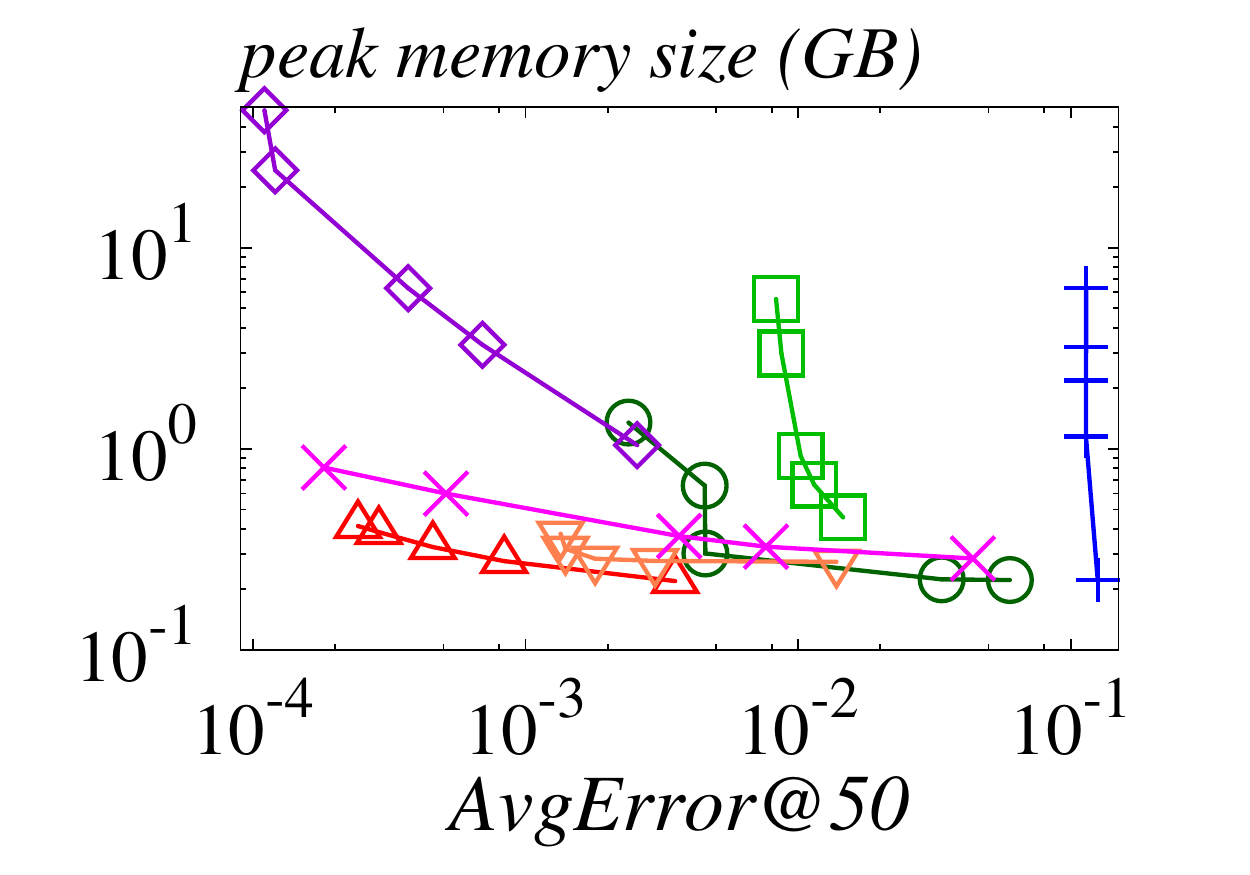}&
\hspace{-5mm} \includegraphics[height=32mm]{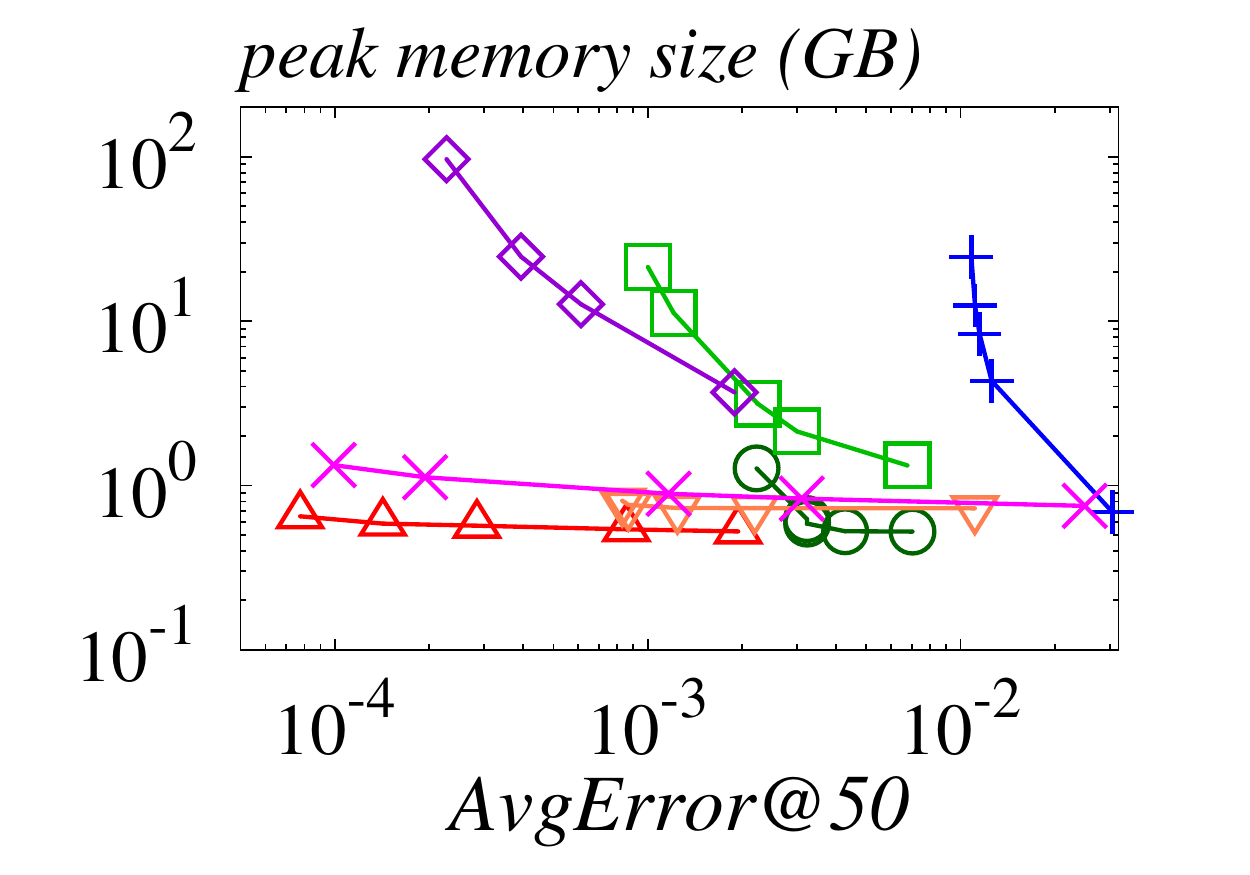} &
\hspace{-5mm} \includegraphics[height=32mm]{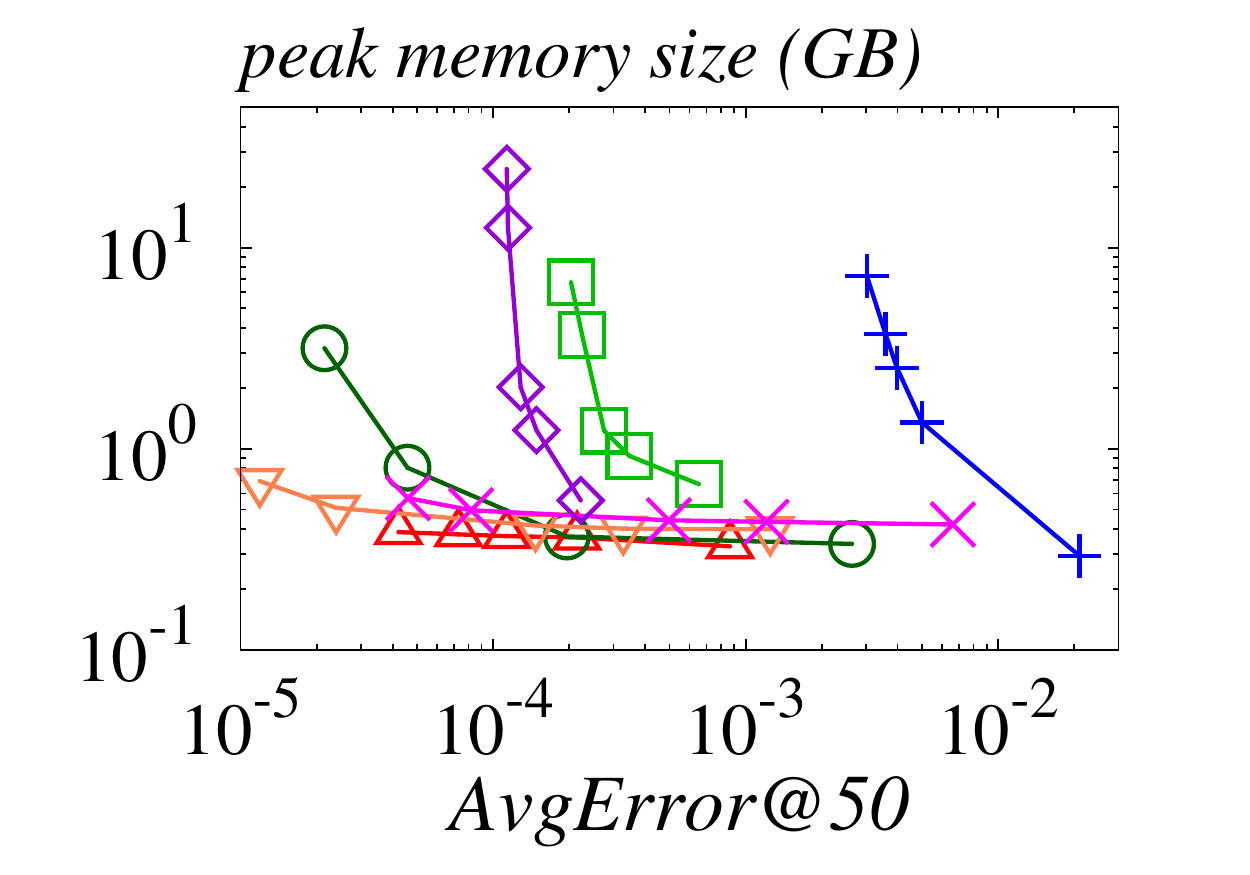}&
\hspace{-5mm} \includegraphics[height=32mm]{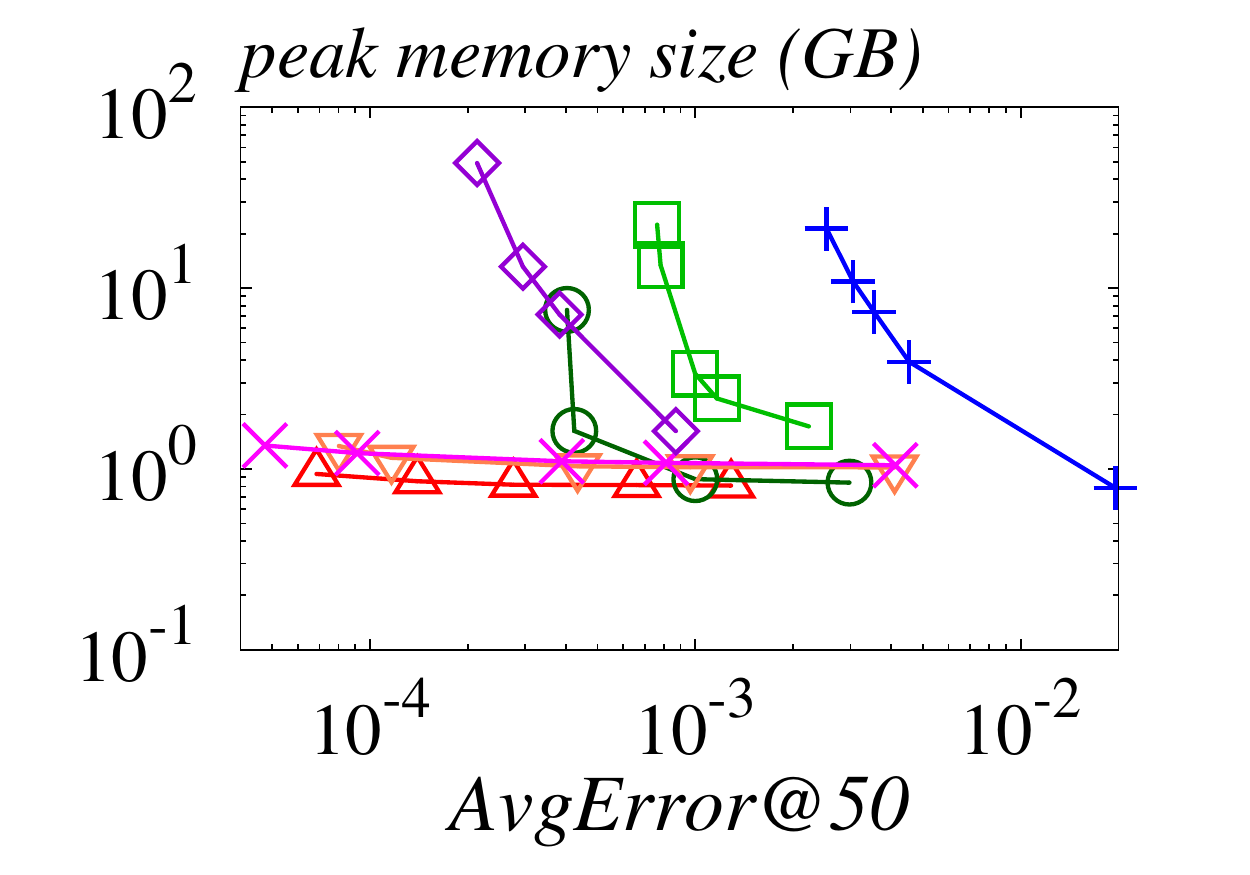} 
\\[-1mm]
\hspace{-4mm} \revise{(a)  In-2004}&
\hspace{-4mm} \revise{(b) DBLP}  &
\hspace{-4mm} \revise{(c) Pokec} &
\hspace{-4mm} \revise{(d) LiveJournal} 
\\[1mm]
 \hspace{-4mm} \includegraphics[height=31mm]{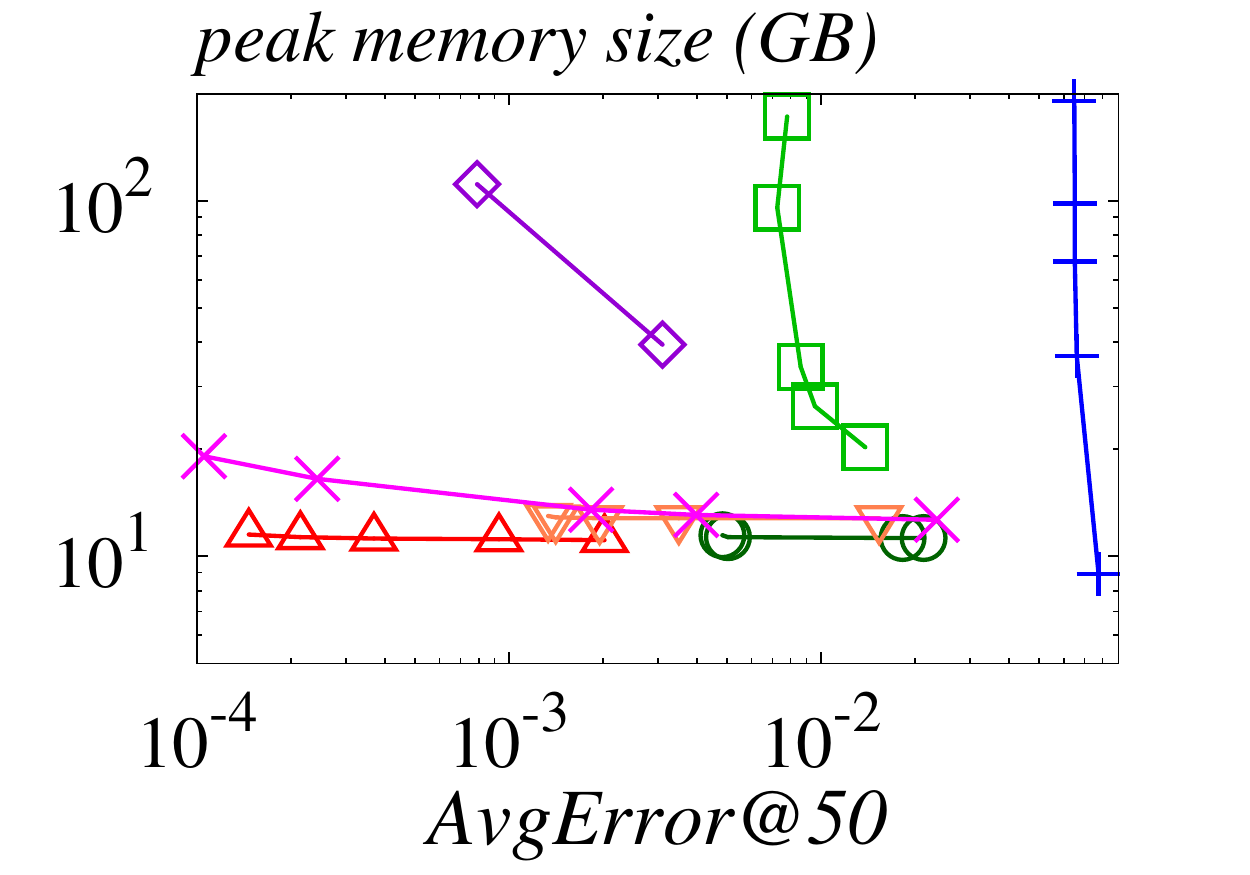} &
\hspace{-4mm} \includegraphics[height=31mm]{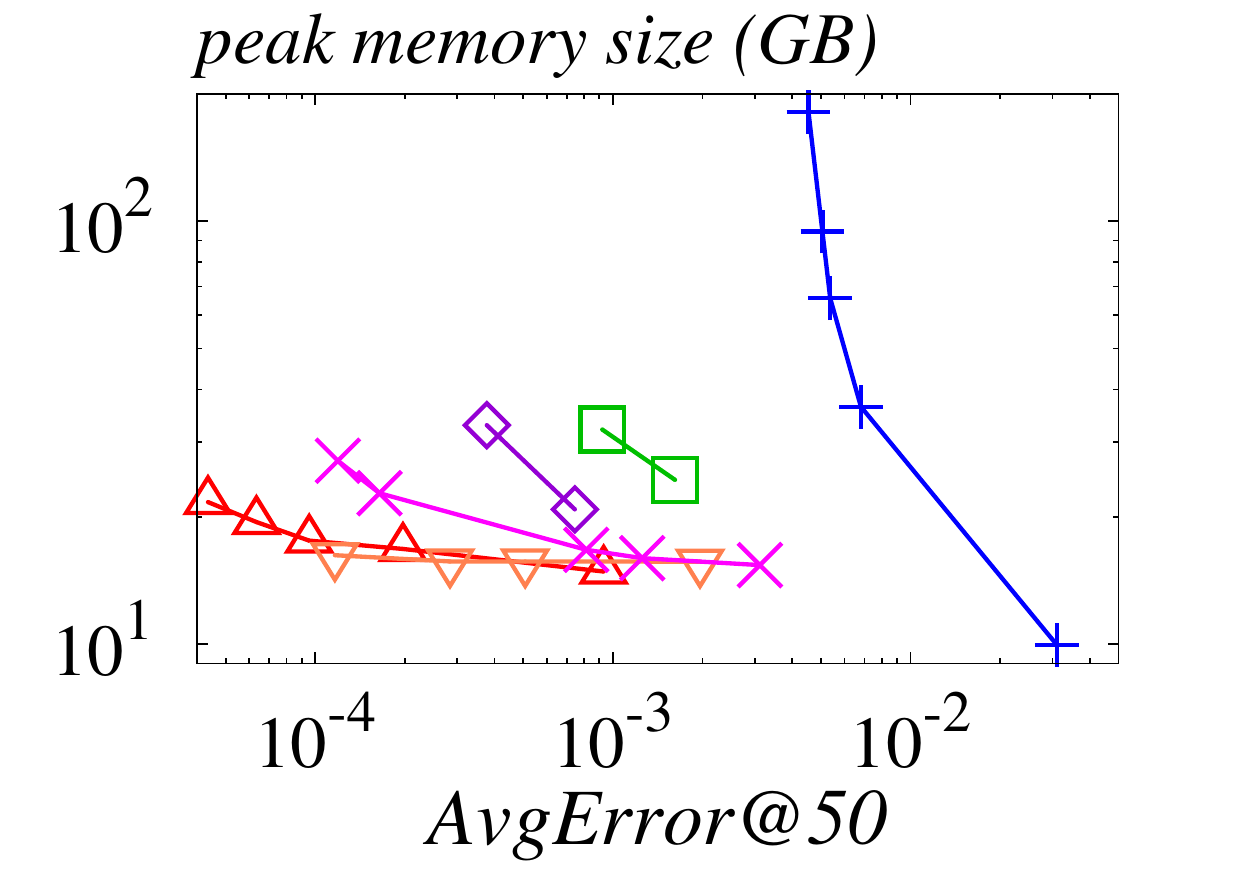}&
\hspace{-4mm} \includegraphics[height=31mm]{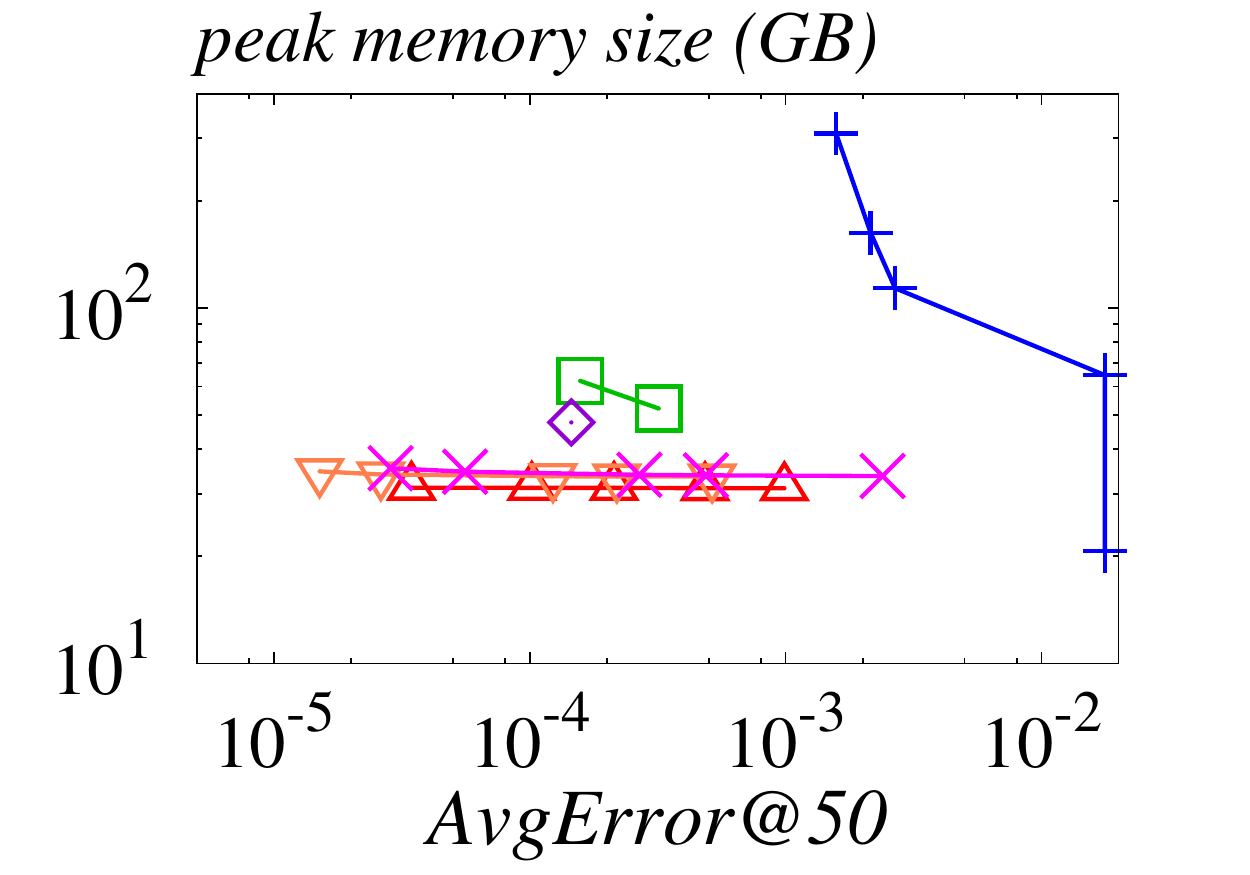} &
\hspace{-4mm} \includegraphics[height=31mm]{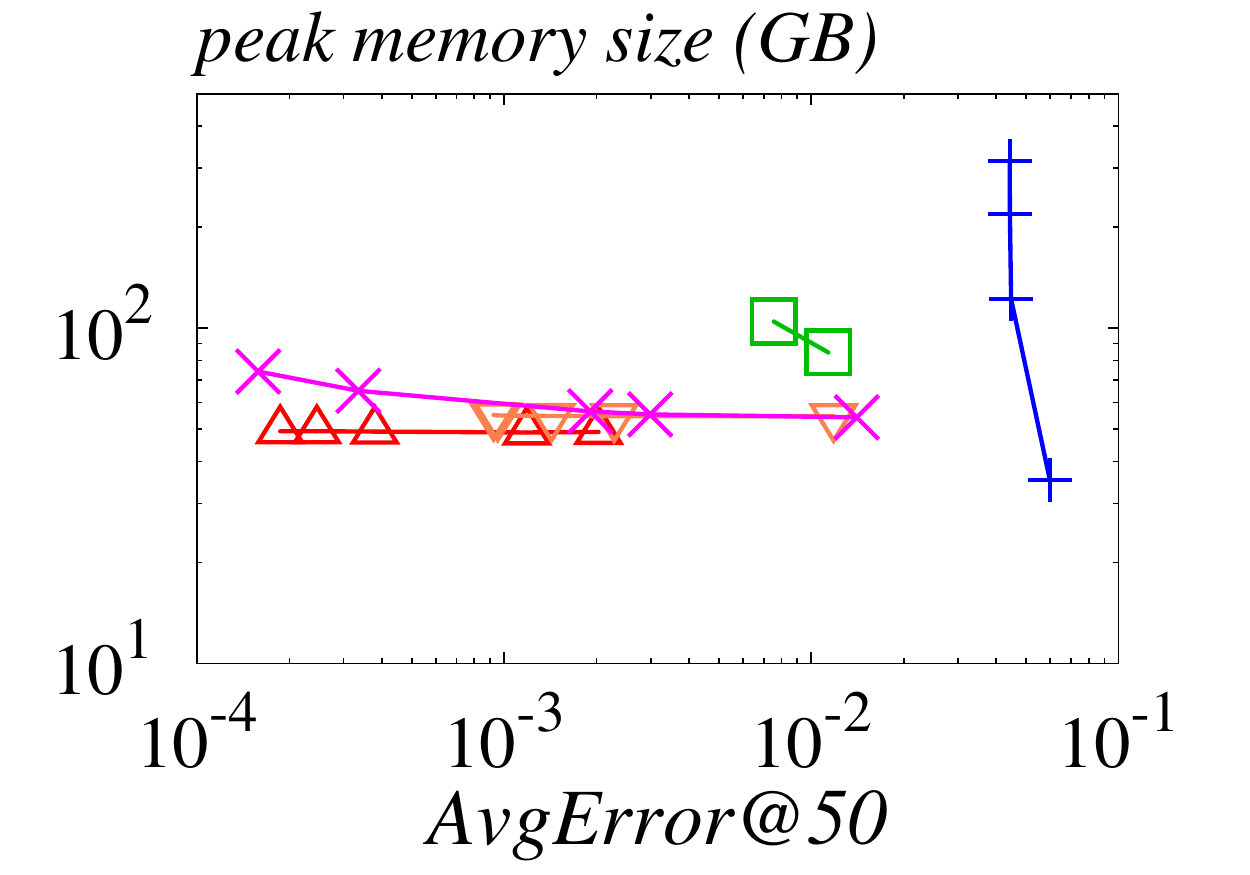}
\\[-1mm]
\hspace{-4mm} \revise{(e) IT-2004}  &
\hspace{-4mm} \revise{(f) Twitter} &
\hspace{-4mm} \revise{(g) Friendster} &
\hspace{-4mm} \revise{(h) UK} \\[-1mm]
\end{tabular}
\vspace{-2.5mm}
\caption{\revise{Average error {\it vs.} peak memory usage.}} \label{exp:err-index}
\vspace{-4mm}
\end{small}
\end{figure*}

\header
\textbf{Average precision and query time.}
Figure \ref{exp:prec-runtime} reports the tradeoff between \textit{Precision@50} and query time of all methods over the first eight graphs \revise{(the evaluation on ClueWeb is presented later)}.
$x$-axis is  {\em Precision@50}, and $y$-axis is  query time in seconds (s) and is in log-scale.
\revise{For each method, the plot contains a curve with 5 points corresponding to its 5 parameter settings (from left to right).}
The overall observation is that \ours provides the best precision and query time tradeoff in most settings, especially on large graphs.
On the large graphs in Figures \ref{exp:prec-runtime}(e)-(h), 
to achieve the same level of precision, \ours is much faster than all index-free and index-based competitors.
For instance, on UK in Figure \ref{exp:prec-runtime}(h), \revise{\ours achieves $96\%$ precision in $0.062$s, while both \probesim and \prsim requires $0.6$s to achieve $96\%$ precision.}
The performance gap between \ours and the competitors remains for varying parameters.
As analyzed, our method \ours focuses computation only on the attention nodes of query $u$, and only explores the vicinity of $u$ to  estimate SimRank values, which is highly efficient. For small graphs in Figures \ref{exp:prec-runtime}(a)-(d), to achieve the same level of precision, {\it e.g.}, above $96\%$,  \tsf, \topsim, \reads, and \sling, are consistently outperformed by \ours.

\header
\textbf{Average error and peak memory usage.}
Figure~\ref{exp:err-index} shows the peak memory usage of all methods. 
The memory usage includes the size of the input graphs, the indices (if any), and any other structures required by the methods.
The $x$-axis is \emph{AvgError@50} and is in log-scale, and the $y$-axis is the peak memory usage in GigaBytes (GB).
\revise{For each method, the  plot contains a curve with 5 points corresponding to its 5 parameter settings, from right to left.}
We find that (i) the peak memory usage of \ours is lower than all competitors over all datasets under almost all settings; (ii) the peak memory usage of \ours is insensitive to $\eps$. The reason is that when decreasing $\eps$, the size of $\Tsrc$ and the number of attention nodes increase  slowly, and thus, \ours can maintain relatively stable peak memory usage.
For instance, in Figure \ref{exp:err-index}(h) for UK graph, \ours requires $48$ to $49$ GB memory, while \probesim needs about $54$ GB and \prsim needs $54$ to $74$ GB.
Methods \sling, \reads, \tsf require much more memory and  are sensitive to parameters.

\begin{figure*}[!t]
\centering
\begin{small}
\begin{tabular}{ccc}
\multicolumn{3}{c}{\hspace{-2mm} \includegraphics[height=2.8mm]{./figure/algo-legend.pdf}}  \\[-1mm]
\hspace{-4mm} \includegraphics[height=30mm]{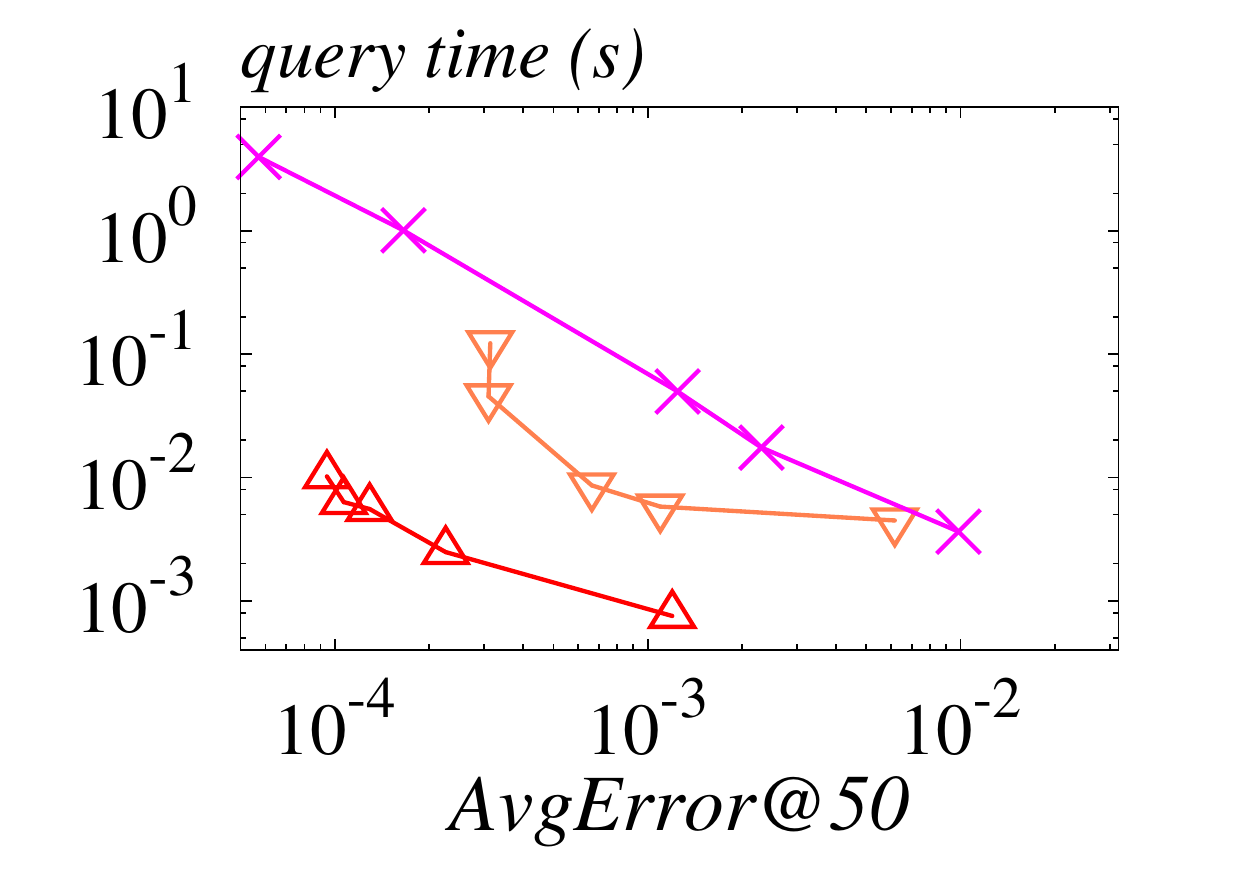}&
\hspace{-4mm} \includegraphics[height=30mm]{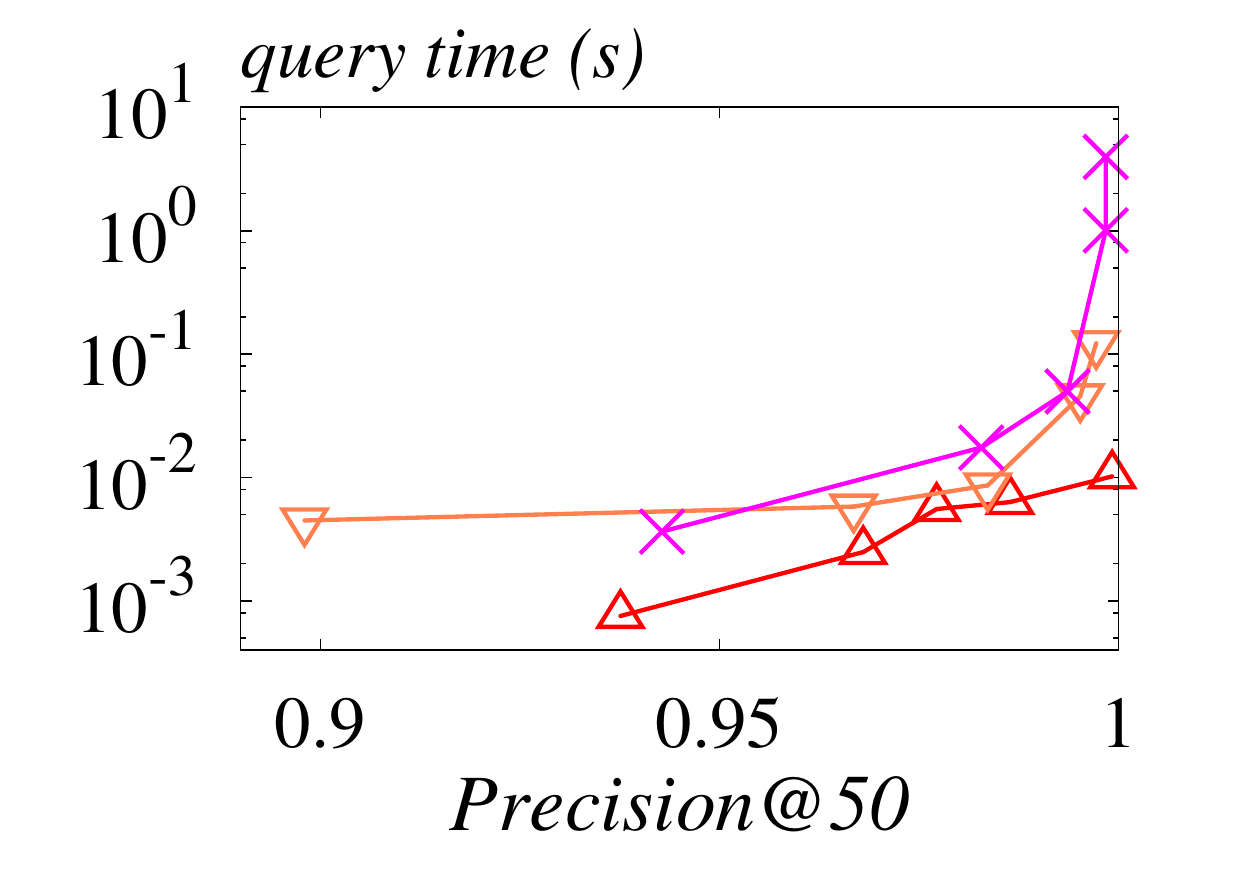} &
\hspace{-4mm} \includegraphics[height=29mm]{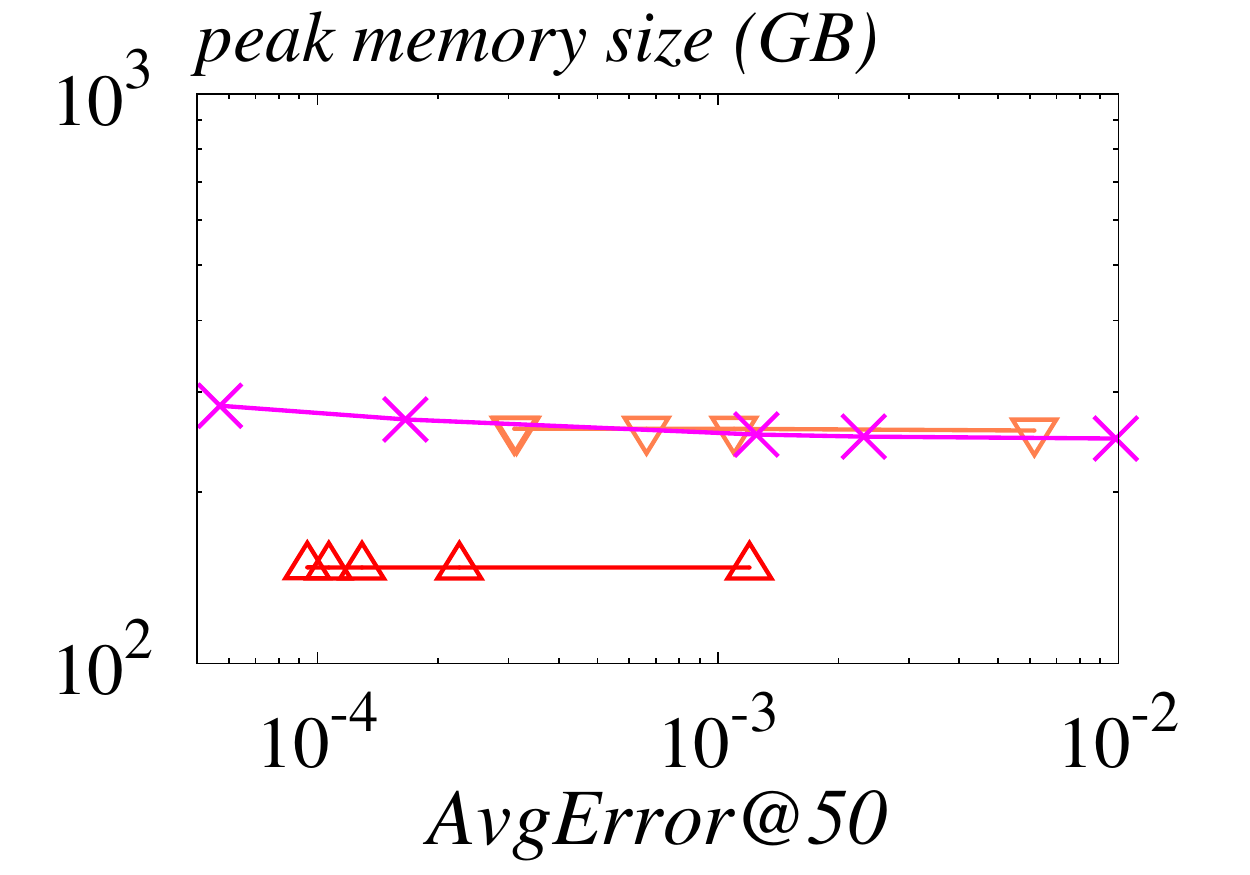}
\\[-1.5mm]
\hspace{-4mm} \revise{(a)} &
\hspace{-4mm} \revise{(b)} &
\hspace{-4mm} \revise{(c)}  
\end{tabular}
\vspace{-3mm}
\caption{\revise{Evaluation on Billion-Node Clueweb}} \label{exp:evaluate-clueweb}
\vspace{-5.5mm}
\end{small}
\end{figure*}

\header
\revise{\textbf{Results on Billion-Node ClueWeb.}
Figure \ref{exp:evaluate-clueweb} reports the evaluation results on the ClubWeb dataset, including \emph{AvgError@50}, \emph{Precision@50}, and peak memory usage measurements.
\tsf, \topsim, \reads, and \sling are not reported since their memory requirements exceed that of our server (376GB).
Figure \ref{exp:evaluate-clueweb}(a) reports the tradeoff between \textit{AvgError@50} and query time. \ours significantly outperforms \prsim and \probesim, often by orders of magnitude. 
Similarly, in Figure \ref{exp:evaluate-clueweb}(b), \ours achieves far more favorable tradeoff between  \textit{Precision@50} and query time.
For instance, to achieve $99.8\%$ precision, \ours takes 0.01\textit{s}, while \prsim needs 1\textit{s} and \probesim uses 0.122\textit{s}.
Figure \ref{exp:evaluate-clueweb}(c) shows the tradeoff between peak memory usage and accuracy. \ours uses about 147 GB memory, whereas \prsim and \probesim each consumes more than 250 GB memory.}

\vspace{-1mm}
\section{Related Work}\label{sec:relatedwork}
We review existing work for SimRank computation, excluding \sling~\cite{slingTianX16}, \probesim\cite{probsimLiuZHWXZL17}, \reads~\cite{readsJiangFWW17}, \tsf~\cite{tsfShaoC0LX15}, \topsim~\cite{topsimLeeLY12} and \prsim~\cite{prsimWeiHX0LDW19}, discussed in Section \ref{sec:stateoftheart}.

Power method \cite{jeh2002simrank} is the first  for all-pair SimRank computation and it computes  SimRank values of all node pairs in the input graph $G$ by the  matrix formulation in \cite{kusumoto2014scalable}:
\begin{equation}\label{eq:power}
\mathbf{S}=(c\mathbf{P}^{\top}\mathbf{S}\mathbf{P})\vee \mathbf{I},    
\end{equation}
where $\mathbf{S}$ is an $n\times n$ matrix such that $\mathbf{S}[i,j]$ is the SimRank value between the $i$-th and $j$-th nodes, $\vee$ is the element-wise maximum operator, $\mathbf{P}$ and $\mathbf{I}$ are the transition matrix and identity matrix of the input graph $G$. Power method starts with 
$\mathbf{S}=\mathbf{I}$, and then updates $\mathbf{S}$ iteratively based on
Equation \eqref{eq:power}, until all elements in $\mathbf{S}$ converge. 
Subsequent studies \cite{lizorkin2008accuracy,yu2015gauging,yu2012space,wang2018efficient} improve the Power method in terms of efficiency or accuracy. However, all these methods incur $O(n^2)$ space overhead, which is prohibitively expensive for web-scale graphs. It is not straightforward to directly apply these methods for single-source SimRank queries.
There are studies \cite{fujiwara2013efficient,he2010parallel,kusumoto2014scalable,li2010fast,yu2014fast,yu2013more,yu2015efficient} that attempt to address the inefficiency issue caused by the operator $\vee$ in Equation \eqref{eq:power}, via an alternative formula for SimRank:
\begin{equation}\label{eq:power2}
\mathbf{S}=c\mathbf{P}^{\top}\mathbf{S}\mathbf{P}+(1-c)\cdot\mathbf{I}.   
\end{equation}
However, as pointed out by \cite{kusumoto2014scalable}, the SimRank computed by Equation \eqref{eq:power2} are rather different from the correct values.

An early work \cite{fogaras2005towards} proposes a Monte Carlo approach to approximate SimRank by sampling conventional random walks. An index structure is also proposed to store random walks. However, the index incurs tremendous space and preprocessing overheads, which makes the Monte Carlo method inapplicable on sizable graphs \cite{slingTianX16,kusumoto2014scalable}.
Maehara \textit{et al.} \cite{MaeharaKK14} propose an index structure for top-$k$ SimRank queries, relying on heuristic assumptions about graphs, and thus, does not provide worst-case error guarantee \cite{probsimLiuZHWXZL17,prsimWeiHX0LDW19}.  
A distributed version of the Monte Carlo approach is proposed by Li et al. \cite{li2015walking}, and the distributed method can scale to a billion-node graph at the significant cost of computation resources; the distributed environment is a different setting that is orthogonal to our study.
There are also studies \cite{antonellis2008simrank,fogaras2005scaling,yu2015high,LinLK12,ZhaoHS09} on variants of SimRank, and SimRank similarity join \cite{MaeharaKK15,TaoYL14}. However, these  solutions are inapplicable for single-source SimRank queries.

\vspace{-1mm}
\section{Conclusion}\label{sec:conclusion}
In this paper, we propose \ours, a novel index-free algorithm that answers single source SimRank queries with rigorous  guarantees, and the method is significantly faster than even the fastest known index-based solutions, often by over an order of magnitude, which is confirmed by our extensive experimental evaluation on real-world web-scale graphs. 
\revise{In the future, we plan to study SimRank queries with relative error guarantees, batch SimRank processing, as well as SimRank computation on new hardware.}

\section*{APPENDIX}

\begin{lemma}
\label{lemlit}
 $\sum_i X_i=N$ and for all $i$, $X_i\in (0,\eps_h)$. $\sum_i Y_i=M$, $M\in (0,1)$ and for all $i$,  $Y_i>0$.  Then $\sum_{i} X_i Y_i\leq \epsp M$.
\end{lemma}
PROOF:
$\sum_{i} X_i Y_i\leq \max_i X_i \cdot \sum_i {Y_i} \leq \epsp M$
\qed

\noindent\textbf{Proof of Lemma \ref{lemma:simoursS1Aset}}.
Obviously, $\simapleft(u,v) \leq s(u,v)$ holds. Now we prove $s(u,v)-\frac{\sqrtc\cdot\epsp}{1-\sqrtc}\leq \simapleft(u,v)$. Let $\Tdst^{(\ell)}$ be the set of all nodes at $\ell$-th level of $\Tdst$.
$\Tdst$ is source graph of $v$ by pushing $L^*$ levels from $v$.
Sum $\sum_{w\in \Tdst^{(\ell)}}h^{(\ell)}(v,w)=\sqrt{c}^{\ell}$. The error  of non-attention nodes at $\ell$-th level:
$\sum_{w\in \Tsrc^{(\ell)}\setminus A_u^{(\ell)} }h^{(\ell)}(v,w) \leq \sqrt{c}^{\ell}.$
For $w\in  \Tsrc^{(\ell)}\setminus A_u^{(\ell)}$,  we have $h^{(\ell)}(u,w)\leq \eps_h$.
Apply Lemma~\ref{lemlit} and $\eta(w)\leq 1$, we have $\textstyle
    \sum_{w\in \Tsrc^{(\ell)}\setminus A_u^{(\ell)} }h^{(\ell)}(v,w)h^{(\ell)}(u,w)\eta(w)\leq \eps_h \sqrt{c}^{\ell}.$
Summing the error of all levels,  $\sum_{\ell=1}  \eps_h \sqrt{c}^{\ell} \leq \frac{\sqrt{c}\eps_h}{1-\sqrt{c}}$.
From Eq.~\eqref{eq:simsling}, 
$    \sum_{\ell=1}\sum_{w\in A_v^{(\ell)} }\lml(u,v,w)=\sum_{\ell=1}\sum_{w\in A_v^{(\ell)}}\hl(u,w)\cdot\eta(w)\cdot\hl(v,w).
$
Thus, $s(u,v)-\frac{\sqrtc\cdot\epsp}{1-\sqrtc}\leq \simapleft(u,v)$. \qed

\noindent
\textbf{Proof of Lemma \ref{lemma:AsetandL}.} 
At level $\ell$, $\sum_{w\in G_u^{(\ell)}}h^{(\ell)}(u,w)=\sqrt{c}^{\ell}$. Hence, at level $\ell$,  there exists at most $\lfloor\frac{\sqrt{c}^{\ell}}{\epsp}\rfloor$ attention nodes, and for $\ell> L^*$, $w\in G_u^{\ell}$, $h^{(\ell)}(u,w)\leq \eps_h$. Therefore, the size of attention set $\Aset$ is at most $\sum_{\ell=1}\lfloor\frac{\sqrt{c}^{\ell}}{\epsp}\rfloor \leq \lfloor \frac{\sqrtc}{(1-\sqrtc)\cdot\epsp}\rfloor$.  \qed

\noindent
\textbf{Proof of Lemma \ref{lemma:simoursS1gamma}.}
$f^{(\ell)}(u,v,w)$ is the $\ell$-$th$ step first meeting probability at $w$, and  we can write $s(u,v)$ as
$
  s(u,v) = \sum_{\ell=1}^{\infty}\sum_{w\in V} f^{(\ell)}(u,v,w).
$
Given $\Asetl$, let
$
    \textstyle
     s_1(u,v)=\sum_{\ell=1}^{L^*}\sum_{w\in \Asetl} f^{(\ell)}(u,v,w),
$
and
$
\textstyle
     s_2(u,v)=\sum_{\ell=1}^{\infty}\sum_{w \notin \Asetl} f^{(\ell)}(u,v,w)
$. Obviously, $s(u,v)=s_1(u,v)+s_2(u,v)$. We want to prove $\simapleftgamma(u,v)\geq s_1(u,v)-s_2(u,v)$ and $s_2(u,v)\leq \frac{\eps_h\sqrt{c}}{1-\sqrt{c}}$.
From Eq.~\eqref{eqdefcor}, 


  \begin{equation}
  \vspace{-3mm}
  \small
  \begin{split}
 \label{eqcor00}
  \simapleftgamma(u,v) 
  = &  \sum\nolimits_{\ell=1}^{L^*}\sum\nolimits_{w\in \Asetl}\hl(u,w)h^{(\ell)}(v,w) \\
 & \times (1- \sum\nolimits_{i=1}^{(L^*-\ell)} \sum\nolimits_{\wli\in \Aset^{(\ell+i)}} \meeti(w,\wli)) \\
  = &  \sum\nolimits_{\ell=1}^{L^*}\sum\nolimits_{w\in \Asetl}[\hl(u,w)h^{(\ell)}(v,w) \\
&  -\sum_{\ell'=1}^{\ell-1}\sum_{w_a \in \Aset^{(\ell')}}{h}^{(\ell')}(u,w_a)h^{(\ell')}(v,w_a)  \rho^{(\ell-\ell')}(w_a,w) ]
\end{split}
 \end{equation}

$f^{(\ell)}(u,v,w)=h^{(\ell)}(u,w)h^{(\ell)}(v,w) 
    -\sum_{\ell':\ell'<\ell}\sum_{w' \in G_u^{(\ell')} }f^{(\ell')}(u,v,w') h^{(\ell')}(w',w)^2
$    

Here we only consider $w'\in A_u^{(\ell')}$, i.e.,  $\hat{f}^{(\ell)}(u,v,w)=$
\begin{equation}
\label{eqff}
\small
h^{(\ell)}(u,w)h^{(\ell)}(v,w)  -\sum_{\ell'<\ell}\sum_{w' \in A_u^{(\ell')} }f^{(\ell')}(u,v,w') h^{(\ell')}(w',w)^2
\end{equation}
 Consider the probability that two $\sqrt{c}$-walks from $u$ and $v$,  first meet at attention node $w'$ then meet at attention node $w$. Given two events: (i) two $\sqrt{c}$-walks from $u$ and $v$ respectively, first meet at some attention node, then meet at $w$, and (ii) these two walks meet at  attention node $w_a$, then two walks from $w_a$ first meet at $w$, the two events hold one-to-one correspondence. 
 The probability of the first event corresponds to the last line of Eq.~\eqref{eqff} and the latter event event corresponds to the last line Eq.~\eqref{eqcor00}. 
Let $\hat{s}_1(u,v)=\sum_{\ell=1}^{L^*}\sum_{w\in \Asetl} \hat{f}^{(\ell)}(u,v,w)$.
Thus, $\simapleftgamma(u,v)= \hat{s}_1(u,v)$.
${s}_1(u,v)-\hat{s}_1(u,v)$ is the probability that two $\sqrt{c}$ walks first meet at non-attention node, then meet at attention node. Thus, ${s}_1(u,v)-\hat{s}_1(u,v)\leq s_2(u,v)$.

Now we prove $s_2(u,v)\leq (\eps_h\sqrtc)/(1-\sqrt{c})$.  Based on Lemma \ref{lemlit},
$
     s_2(u,v)= \sum\nolimits_{\ell=1}\sum\nolimits_{w\in \Tsrc^{(\ell)}\setminus A_u^{(\ell)}} f^{(\ell)}(u,v,w) 
             \leq   \sum_{\ell=1} \sum\nolimits_{w\in \Tsrc^{(\ell)}\setminus A_u^{(\ell)} }h^{(\ell)}(v,w)h^{(\ell)}(u,w) 
             \leq   \frac{\sqrt{c}\cdot\eps_h}{1-\sqrt{c}},
$
Thus, $s(u,v)\geq \simapleftgamma(u,v)\geq  \hat{s}_1(u,v) \geq s_1(u,v)-s_2(u,v)\geq s(u,v)-2s_2(u,v)\geq s(u,v)-\frac{2\sqrt{c}\eps_h}{1-\sqrt{c}}$. \qed

\noindent 
\textbf{Proof of Lemma \ref{lemma:simoursEpsp}.}
In Algorithm~\ref{alg:revspread},  consider the  lose of Simrank  at level $\ell$. Similar to prove Lemma \ref{lemma:simoursS1Aset}, the lose at level $\ell$ $\leq \eps_h \cdot \sqrt{c}^{\ell}$. Summing up all levels, the   total loss is $\leq \frac{\eps_h\cdot \sqrt{c}}{1-\sqrt{c}}$.  Thus $s(u,v)-\tilde{s}(u,v)\leq \frac{3\eps_h\sqrt{c}}{1-\sqrt{c}}\leq \eps$.  \qed

\noindent
\textbf{Proof of Lemma \ref{lemma:leftpush}.}
We push $O(L^*)=O(\log \frac{1}{\eps})$ levels and each level needs $O(m)$ times, and thus the total time is $O(m \log \frac{1}{\eps})$. 
Let $\hat{h}^{(\ell)}(u,w)$ be the Monte Carlo estimation of ${h}^{(\ell)}(u,w)$.
From Hoeffding bound~\cite{hoeffding1963probability}, 
$
    \Pr(\hat{h}^{(\ell)}(u,w)\geq {h}^{(\ell)}(u,w)-\eps_h/2)
    \geq 1-\exp[-2(\eps_h/2)^2 \cdot 
2\log \frac{1}{(1-\sqrt{c})\eps_h\delta}/\eps_h^2]\geq 1-(1-\sqrt{c})\eps_h\delta.
$
Since attention nodes are at most $\lfloor \frac{\sqrtc}{(1-\sqrtc)\cdot\epsp}\rfloor$, 
applying union bound,
 with probability at least $1-\delta$, $\Tsrc$ contains all nodes $u$ with  ${h}^{(\ell)}(u,w)\geq \eps_h$, for all $\ell$. The expected time of MC is $O(\log \frac{1}{\eps\delta}/\eps^{2})$. \qed

\noindent\textbf{Proof of Lemma \ref{lemma:lastmeetingcomplex}.}
Algorithm \ref{alg:pushintree} costs $O(m)$ per level of $\Tsrc$. Node $w$ has $O(1/\eps)$ hitting probabilities from $w$. Thus the complexity of one level is $O(m/\eps)$. There are $O(\log \frac{1}{\eps})$ levels. Total complexity is $O(m\log \frac{1}{\eps}/\eps)$.
Let $Z_i$ be the number of nodes in $\Tsrc$ at level $i$. 
For all $w_{1}\in \Tsrc^{(\ell+1)}$,  the cost of all $\rho^{(1)}(w,w_1)$ is $O(Z_{l+1})$.  For all $w_{2}\in \Tsrc^{(\ell+2)}$, from Eq.~\eqref{eqdefcor}, the cost of all $\rho^{(2)}(w,w_2)$ is  $O(Z_{\ell+1}Z_{\ell+2})$. Similarly, we can compute all $w_i\in  \Tsrc^{(\ell+i)}$ for all $i>0$ in 
$ \textstyle
     O(Z_{\ell+1}+\sum_{i=1} \sum_{j=1} Z_{\ell+i}Z_{\ell+i+j})
$ time. 
$Z_{\ell+1}+\sum_{i=1} \sum_{j=1} Z_{\ell+i}Z_{\ell+i+j} \leq (\sum_{i=1} Z_i)^2$ and $\sum_{i=1} Z_i\leq O(1/\eps)$, then $O(Z_{\ell+1}+\sum_{i=1} \sum_{j=1} Z_{\ell+i}Z_{\ell+i+j})\leq O(1/\eps^2)$.  Thus  the cost of computing $\corrl(w)$ for all attention nodes $O(\frac{1}{\eps^3})$.
The total complexity is $\max\{ \frac{m\log \eps}{\eps},  \frac{1}{\eps^3} \}$.  \qed

\noindent\textbf{Proof of Lemma \ref{lemma:revspreadcomplex}.}  Algorithm~\ref{alg:revspread} costs $O(m)$ per level and have $O(\log\frac{1}{\eps})$ levels. The total cost is $O(m \log \frac{1}{\eps})$. \qed

\noindent\textbf{Proof of Theorems \ref{theorem:correctness} $\&$ \ref{theorem:timecomplex}.}
Given Lemma \ref{lemma:simoursS1gamma}, \ref{lemma:simoursEpsp}, \ref{lemma:leftpush}, Theorem~\ref{theorem:correctness} follows. Given Lemma~\ref{lemma:leftpush}, \ref{lemma:lastmeetingcomplex}, \ref{lemma:revspreadcomplex}, Theorem~\ref{theorem:timecomplex} follows. \qed


\balance




\bibliographystyle{abbrv}
\bibliography{sigproc}  


\end{document}